\documentclass[12pt]{iopart}

\usepackage{graphicx}% Include figure files
\usepackage{dcolumn}% Align table columns on decimal point
\usepackage{bm}% bold math
\usepackage{amsmath,amsthm,amssymb}
\usepackage{mathrsfs}  
\usepackage{comment}
\usepackage{tikz} 
\usepackage{cite} 
\usepackage{physics}
\usepackage{mathtools}
\usetikzlibrary{arrows.meta,bending,matrix,positioning}
\newenvironment{tbs}{%
   \small\tt
   \begin{enumerate}[$\blacktriangleright$]}{\end{enumerate}}
\newcommand{\btbs}{\begin{tbs}}                                                                      
\newcommand{\etbs}{\end{tbs}}

\newcommand{\Reals}{\mathbb{R}}

\newcommand{\x}{ {\bf x}}
\newcommand{\y}{{\bf y}}
\newcommand{\reals}{\Reals}

 \newtheorem{remark}{Remark}
 \newtheorem{lemma}{Lemma}
 \newtheorem{corollary}{Corollary}
 \newtheorem{theorem}{Theorem}
 \newtheorem{definition}{Definition}
 \newtheorem{proposition}{Proposition}
 \newtheorem{example}{Example}

 \makeatletter
\DeclareRobustCommand\frownotimes{\mathbin{\mathpalette\frown@otimes\relax}}
\newcommand{\frown@otimes}[2]{%
  \vbox{
    \ialign{##\cr
      \hidewidth$\m@th#1{}_\frown$\kern-\scriptspace\hidewidth\cr
      \noalign{\nointerlineskip\kern-1pt}
      $\m@th#1\otimes$\cr
    }%
  }%
}
\makeatother

\begin{document}

%\preprint{APS/123-QED}

\title[]{Why we should interpret density matrices as moment matrices: the case of  (in)distinguishable  particles and the emergence of classical reality}% Force line breaks with \\

\author{Alessio Benavoli}
 \address{School of Computer Science and Statistics, Trinity College Dublin, Ireland}%Lines break automatically or can be forced with \\
 \ead{alessio.benavoli@tcd.ie}
\author{Alessandro Facchini \& Marco Zaffalon}%
 \address{Dalle Molle Institute for Artificial Intelligence Research (IDSIA), Lugano, Switzerland.}
  \ead{alessandro.facchini@idsia.ch, marco.zaffalon@idsia.ch}

\vspace{10pt}
\begin{indented}
\item[]\today
\end{indented}

\begin{abstract}
We introduce a formulation of quantum theory (QT) as  a general   probabilistic   theory   but expressed via   quasi-expectation  operators (QEOs). This formulation  provides  a  direct interpretation  of  density  matrices  as  quasi-moment  matrices.
Using QEOs, we will provide a series of \textit{representation theorems}, {\`a} la de Finetti, relating a classical probability mass function (satisfying certain symmetries) to a quasi-expectation operator. We will show that QT for both distinguishable and indistinguishable particles can be formulated in this way. Although particles indistinguishability is considered a truly  `weird'  quantum  phenomenon,  it  is  not  special. We  will  show  that  finitely  exchangeable probabilities for a classical dice are as weird as  QT.  Using this connection,  we will rederive the first and second quantisation in QT for bosons through the classical statistical concept of exchangeable random variables.  Using  this  approach,  we  will  show  how  classical  reality emerges  in  QT  as  the    number  of  identical bosons increases (similar to what happens for finitely exchangeable sequences of rolls of a classical dice).
\end{abstract}

\maketitle

%\tableofcontents

\section{Introduction}
\label{sec:intro}
General probabilistic theories (GPTs) are a family of operational theories that generalize both finite-dimensional Classical Probability Theory (CPT) and finite-dimensional Quantum Theory (QT) \cite{birkhoff1936logic,mackey2013mathematical,jauch1963can,hardy2011foliable,hardy2001quantum,barrett2007information,chiribella2010probabilistic,barnum2011information,van2005implausible,pawlowski2009information,dakic2009quantum,fuchs2002quantum,brassard2005information,mueller2016information,coecke2012picturing,Caves02,Appleby05a,Appleby05b,Timpson08,longPaper,Fuchs&SchackII,mermin2014physics,pitowsky2003betting,Pitowsky2006,benavoli2016quantum,benavoli2017gleason,popescu1998causality,navascues2010glance}.

There are several approaches to GPTs (see
\cite{janotta2014generalized,plavala2021general} for a review), but they are  either equivalent or only slightly different. GPTs were formalised with the goal of deriving QT from a set of reasonably motivated principles. Moreover, GPTs allow to reformulate QT involving only real-valued vector spaces. Overall operational theories have proven successful in disentangling the differences between CPT and QT. 

 By identifying states with the density matrices, one faces the problem that, whereas the latter live in a complex-valued Hilbert space, the GPT framework only involves real-valued vector spaces. The solution adopted in GPTs to overcome this issue is to exploit the fact that a $n \times n$ density matrix can be parametrised in terms of $SU(n)$  generators with real coefficients. For example, in the qubit ($n=2$) case, the  density matrix can be expressed in terms of $SU(2)$-generators (Pauli matrices) as
$$
\rho = \frac{1}{2}(I_2 + a \sigma_x + b \sigma_y + c \sigma_z),
$$
with real coefficients $a, b, c \in [-1, 1]$ satisfying the constraint $a^2 + b^2 + c^2 \leq 1$.
Although this approach can be extended to any dimension $n>2$, the constraints on the real coefficients become more and more complex at the increase of $n$ \cite{bengtsson2017geometry}. This explains why, despite the success of the GPT program, we are still using the old QT formalism referring to Hilbert spaces.

In this work, we present a different but equivalent way to formulate QT as a GPT, that is in terms of quasi-expectation operators (QEOs). This approach is dual  to the \textit{algorithmic (bounded) rationality theory} we introduced in \cite{Benavoli2021f}. This formulation departs from standard GPT  in three ways. First, it focuses on expectation operators rather than probability measures. Second, the QEO is generally defined on a vector space of real-valued functions (e.g., polynomials) whose underlying variables take values in an infinite dimensional space of possibilities. When the QEO  is an expectation operator, these variables play naturally the role of hidden-variables.  Third, QEOs  become finite dimensional when the (quasi)-expectation operator is restricted to act on a finite-dimensional vector space of real-valued functions (as it is for QT).

Under the QEO framework, we demonstrate that, maybe not surprisingly, density matrices have a  natural interpretation as \textit{quasi-moment matrices}, that is they are similar to the covariance matrix of a Gaussian distribution, which is indeed a positive semi-definite matrix.

Using QEOs, we will provide a series of \textit{representation theorems}, {\`a} la de Finetti, relating a classical probability mass function (satisfying certain symmetries) to a QEO $\widetilde{L}$ defined over a vector space of polynomials:
$$
\textit{probability}= \widetilde{L}(\textit{polynomials}).
$$
We will show that QT for both distinguishable and indistinguishable particles can be formulated in this way.
In particular, we will rederive the \textit{first and second quantisation} in QT    through the classical statistical concept of \textit{exchangeable sequence of random variables}.

Although particles indistinguishability is considered a truly `weird' quantum phenomenon, it is not special. Indeed, we will show in Section \ref{sec:defin} that finitely exchangeable probabilities for a classical dice are as weird as QT. Starting from de Finetti's representation theorem, we discuss a series of representation theorems for the probability of finitely exchangeable rolls of a classical dice. These representation theorems involve negative probabilities (and entanglement) or, equivalently, QEOs similarly to what happens in QT. We will then discuss how the weirdness disappears as the  considered number of rolls of the dice increases.

We will then use the same approach to derive a representation theorem for the second quantisation  for bosons. Using this approach, we will show how classical reality emerges in QT as the considered number of identical bosons increases.

\section{Explaining QEO}
\label{sec:QET}
A CPT is usually stated in terms of probability axioms. However, it can be more generally formulated from axioms on the expectation operator \cite{whittle2000probability}. 

Consider a vector  of variables ${\bf x}$ taking value in the possibility space $\Omega$, and a vector space $\mathcal{F}$ of real-valued bounded functions  on ${\bf x}$ including the constants.

\begin{definition}[{\cite[Sec.\ 2.8.4]{walley1991}}]\label{def:expect}
 Let $L$ be a  linear functional $L: \mathcal{F}\rightarrow \reals$. $L$ is an \textbf{expectation operator}  if it satisfies the following property: 
\begin{equation}
\label{eq:A}
     L(g)\geq \sup c \text{ s.t. } g-c \in \mathcal{F}^+,  \tag{A} 
\end{equation}
for every $g \in \mathcal{F}$, where $\mathcal{F}^+$ is the closed convex cone\footnote{A subset $\mathcal{C}$ of a real-vector space $\mathcal{F}$ is a cone if for each $f \in \mathcal{F}$ and positive scalar $\alpha>0$, the element  $\alpha f$ is in $\mathcal{C}$. A cone $\mathcal{C}$ is a convex cone if $\alpha f +\beta g$  belongs to $\mathcal{C}$, for any scalars  $\alpha,\beta>0$ and $f,g \in \mathcal{F}$.} of nonnegative functions in $\mathcal{F}$ and $c$ is the constant function of value $c$. 
\end{definition}
It can be easily verified that \eqref{eq:A} is equivalent to:
\begin{equation}
\label{eq:Ainf}
     L(g)\geq \inf_{{\bf x} \in \Omega} g({\bf x}).
\end{equation}
In the sequel, to simplify the notation, we simply write $\inf g$ instead of $\inf_{{\bf x} \in \Omega} g({\bf x})$. Linearity and \eqref{eq:Ainf} are the two properties that define a classical expectation operator. Indeed, from these two properties, we can derive that
\begin{itemize}
 \item $L(0)=0$;
 \item $0\stackrel{A}{=}L(0)=L(g-g)\stackrel{linearity}{=}L(g)+L(-g)$ and so $L(g)=-L(-g)$;
\end{itemize}
which, together with $L(-g)\geq \inf -g=-\sup g$, leads to
\begin{equation}
\label{eq:infsupbound}
 \inf g \leq L(g) \leq \sup g.
\end{equation}
This means that $L(g)$ is a `weighted-average': the weights being the probability measure associated to the expectation operator; note in fact that $\inf g \leq \int_{\Omega} g\, dp \leq \sup g$ for any probability measure $p$. This formulation in terms of probabilities is not necessary. Indeed, we can more generally work with expectation operators.

A quasi-expectation operator is a conservative relaxation of an expectation operator. It is defined as follows.

\begin{definition}\label{def:TET}
 Let $\widetilde{L}$ be a linear functional $\widetilde{L}: \mathcal{F}\rightarrow \reals$ and $\mathcal{C}^+$ be a closed convex cone (including the constants) such that $\mathcal{C}^+\subseteq \mathcal{F}^+$. We call $\tilde{L}$ a \textbf{quasi-expectation} operator (QEO)  if it satisfies
 \begin{equation}
 \label{eq:Astar}
   \widetilde{L}(g)\geq \sup c \text{ s.t. } g-c \in \mathcal{C}^+, \tag{A$^*$}   
 \end{equation}
 for every $g \in \mathcal{F}$.
A QEO is called (computationally) \textbf{tractable}, whenever  the membership $g-c \in \mathcal{C}^+$ can be computed in \textbf{P}-time.
\end{definition}
Let $\underline{c}_g$ be equal to the supremum value of $c$ such that $g-c \in \mathcal{C}^+$. It can be verified that \eqref{eq:Astar} implies:
 \begin{equation}
 \label{eq:inequcg}
   \widetilde{L}(g)\geq \underline{c}_g, \text{  where}   \inf_{{\bf x} \in \Omega} g({\bf x})\geq \underline{c}_g.   
 \end{equation}
 
A QEO (conservatively) relaxes  property \eqref{eq:A} by providing a lower bound  $\underline{c}_g$ to $\inf  g$.

From property \eqref{eq:Astar}, similarly to what was done for expectation operators, we can derive
\begin{equation}
\label{eq:infsupboundgen}
\underline{c}_g \leq \widetilde{L}(g) \leq \overline{c}_g,
\end{equation}
with $\overline{c}_g=-\inf -g$, where 
\begin{equation}
\label{eq:infsupboundgen1}
\underline{c}_g  \leq \inf g  \leq \sup g \leq \overline{c}_g.
\end{equation}
Notice that since the external inequalities of Equation~\eqref{eq:infsupboundgen1} can be strict for some $g$, we cannot in general define  $\widetilde{L}$ as an integral with respect to a probability measure and, therefore,  $\widetilde{L}(g)$ cannot be a `weighted average'. In other words, $\widetilde{L}$ is not a classical expectation operator. In general, in order to  write $\widetilde{L}(g)$ as  an integral and satisfy \eqref{eq:infsupboundgen1}, we need to introduce some negative values:
$$
\widetilde{L}(g)=\int_{\Omega} g\, d\nu,
$$
where $\nu$ is a signed-measure. As we proved in the \textit{Weirdness Theorem} in \cite[Th.\ 1]{Benavoli2021f}, the condition \eqref{eq:Astar} characterises the condition under which
the weirdness shows up in a  theory. In general, any theory satisfying \eqref{eq:Astar} with $ \mathcal{C}^+\subset \mathcal{F}^+$ will have negative probabilities and non-classical (non-boolean) evaluations functions.

To sum up, under the considered formalism, a GPT is thus obtained by providing variables ${\bf x}$ taking value in a space of possibility $\Omega$, a vector space $\mathcal{F}$ of real-valued bounded functions  on ${\bf x}$ including the constants, and finally a QEO $\widetilde{L}$ over $\mathcal{F}$. 

Whenever $\widetilde{L}$ is a tractable QEO, we also call the corresponding GPT tractable. In this case, we can interpret the  GPT as an \textit{algorithmic (bounded) rationality theory} \cite{Benavoli2021f}.  These tractable theories model the idea that rationality (expressed by the property \eqref{eq:A} in CPT) is limited by the available computationally resources for decision making. In this case, for the decision-maker, it may only be possible to impose a weaker form of rationality \eqref{eq:Astar}, but that can be efficiently computed. 
As explained in the next section, QT is an instance of such tractable theories.

%\begin{equation}
%\label{eq:TETbounds}
%\underline{c}_g \leq \inf g  \leq \sup g \leq \overline{c}_g,
%\end{equation}

\subsection{Quantum Theory via QEOs}
Let us now move to the QT setting. In order to provide our representation theorem for QT in terms of QEOs, we  need to define $({\bf x},\mathcal{F},\widetilde{L})$.
In QT, the unknown variable is  ${\bf x} \in \Omega= \overline{\mathbb{C}}^{n_x}$ with
\begin{equation}
\label{eq:Cbar}
    \overline{\mathbb{C}}^{n_x}:=\{{\bf x} \in \mathbb{C}^{n_x}:~~{\bf x}^{\dagger}{\bf x}=1\},
\end{equation}
and the vector space of real-valued bounded functions 
\begin{equation}
 \label{eq:gambF1}
\mathcal{F}=\{g({\bf x},{\bf x}^\dagger)={\bf x}^\dagger G {\bf x} : G \text{ is a Hermitan matrix}\},
\end{equation}
which includes the constants (take $G=cI_{n_x}$ where $I_{n_x}$ is the identity matrix of dimension $n_x$ then ${\bf x}^\dagger cI {\bf x}=c$ for any $c \in \mathbb{R}$).  In QT, $\mathcal{F}$ is the set of observables for a single-particle system with $n_x$ degrees of freedom.  $\mathcal{F}$ includes functions whose we can compute expectations by performing an experiment. In QT, observables are usually denoted as Hermitian operators $G$, but $G$ is not a function. $G$ includes the coefficients of the quadratic form ${\bf x}^\dagger G {\bf x}$.
%, which is the function whose we aim to compute the expectation.

Since in QT it refers to the average value of the observable represented by operator $G$ for the physical system in the state $\ket{\x}$, in our work we do not use the notation $\expval{G}{\x}$. As in CPT, $\x$ is used to denote an unknown variable and  ${\bf x}^\dagger G {\bf x}$  is the quantity of which we are interested in calculating the expectations. %We will further clarify this difference later on.

Observe that in \eqref{eq:gambF1}, we write the function $g$ as $g({\bf x},{\bf x}^\dagger)$ and not as $g({\bf x})$, because a complex number $z$ and its conjugate $z^\dagger$ are effectively different numbers (contrasted to $a$ and $a^\top$ for $a \in \mathbb{R}$). This difference gives rise to many of the properties of QT.

\begin{theorem}[Representation theorem for one-particle systems]
\label{th:reprsingle}
For every   $g({\bf x},{\bf x}^\dagger)={\bf x}^\dagger G {\bf x} \in \mathcal{F}$  in \eqref{eq:gambF1}, the following definitions are equivalent
\begin{enumerate}
\item ${L}:\mathcal{F}\rightarrow \reals$ is a valid expectation operator, that is it satisfies property \eqref{eq:A}.
\item $L$ can be written as 
\begin{equation}
 \label{eq:expM0}
L(g)= Tr(GM),
\end{equation}
where $M$ is a $n_x \times n_x$ Hermitian matrix such that $M\succeq 0$ (PSD) and $Tr(M)=1$.
\end{enumerate}
\end{theorem}
All the proofs are in Supplementary \ref{sec:proofs}.
Note that $L: \mathcal{F}\rightarrow \reals$ is a real-valued operator defined on the space of real-valued functions $\mathcal{F}$. Therefore, as in GPT, QEO is defined on a real-vector space.

In Theorem~\ref{th:reprsingle}, $L(g)$  is a tractable expectation operator as stated in the following well-known result.
\begin{proposition}\label{prop:eig}
The infimum (minimum) of $g({\bf x},{\bf x}^\dagger)={\bf x}^\dagger G{\bf x}$ is equal to the minimum eigenvalue of $G$, and can therefore computed in \textbf{P}-time. 
\end{proposition}

In Theorem~\ref{th:reprsingle}, $L(g)$  is a classical expectation.
Indeed, for one-particle systems, QT is compatible with CPT.
To explain the relation between $L(g)$ and CPT (probability measures), consider a probability distribution $p$ on the unknown $\x$. The expectation of ${\bf x}^\dagger G {\bf x}$ w.r.t.\ $p$ is then given by:
\begin{equation}
 \label{eq:momM}
 \begin{aligned}
&\int_{\Omega} {\bf x}^\dagger G {\bf x}\, p({\bf x}) d{\bf x}=\int_{\Omega} Tr(G  {\bf x}{\bf x}^\dagger) p({\bf x}) d{\bf x}\\
  &=Tr\left(G \int_{\Omega} {\bf x}{\bf x}^\dagger p({\bf x}) d{\bf x}\right)=Tr(GM),
 \end{aligned}
\end{equation}
where we have exploited the linearity of the trace and expectation, and defined the matrix
\begin{equation}
 \label{eq:momMdefined}
M:=L({\bf x}{\bf x}^\dagger)=\int_{\Omega} {\bf x}{\bf x}^\dagger p({\bf x}) d{\bf x}.
\end{equation}
For the formal derivation of \eqref{eq:momMdefined}, we extended $L: \mathbb{C}[\x]\rightarrow \mathbb{C}$, where $\mathbb{C}[\x]$ is the polynomial ring in $\x$ over $\mathbb{C}$, then the expectation operator $L$ is applied element-wise to ${\bf x}{\bf x}^\dagger$. 
\begin{example}
For instance, for one-particle system with $n_x=3$ and $\x=[x_1,x_2,x_3]^\top$, we have
\begin{equation}
L({\bf x}{\bf x}^\dagger)=
\begin{bmatrix}
L(x_1x_1^\dagger)  & L(x_1x_2^\dagger) & L(x_1x_3^\dagger)\\ 
L(x_2x_1^\dagger)  & L(x_2x_2^\dagger) & L(x_2x_3^\dagger)\\ 
L(x_3x_1^\dagger)  & L(x_2x_2^\dagger) & L(x_3x_3^\dagger)\\ 
    \end{bmatrix}.
\end{equation}
Note that, $L(x_1x_1^\dagger)+L(x_2x_2^\dagger)+L(x_3x_3^\dagger)=L({\bf x}^\dagger \x)=1$.
\end{example}

Theorem~\ref{th:reprsingle} implies that the set of \textbf{belief states}, called density matrices in QT,  is
$$
\{M \text{ is a $n_x \times n_x$ Hermitian matrix}: M\succeq 0, ~~Tr(M)=1\}.
$$
Because of \eqref{eq:momMdefined}, we can interpret a belief state (aka a density matrix) as a (truncated) moment matrix,  similar to the covariance matrix of a Gaussian distribution. This also implies that we do not need a probability distribution to define an expectation operator. Indeed, in general, infinitely many probability measures have $M$ has truncated moment matrix. For instance, if $p_1({\bf x}),p_2({\bf x})$ have $M$ as moment matrix, then $p_3({\bf x})=\alpha p_1({\bf x})+(1-\alpha )p_2({\bf x})$ for any $\alpha \in (0,1)$ has $M$ as moment matrix. This is related to the \textit{preferential basis problem} in QT, which simply follows by the fact that expectation operators are more general than probabilities.

To sum up, Theorem~\ref{th:reprsingle} tells us that, in the case of a single particle system, QT is an instance of a tractable GPT compatible with CPT. % since full rationality is tantamount to algorithmic rationality, QT and CPT are compatible.
The situation is different for two (or more) particles' systems.

Consider another particle ${\bf y} \in \overline{\mathbb{C}}^{n_y}$ and the vector space of real-valued bounded functions 
\begin{equation}
 \label{eq:gambF2}
\mathcal{H}:=\{h({\bf y},{\bf y}^\dagger)={\bf y}^\dagger H {\bf y} : H \text{ is a Hermitan matrix}\}.
\end{equation}
Independence judgements between the variables ${\bf x},{\bf y}$ can be expressed in terms of expectations by stating 
$$
L(gh)=L(g)L(h),$$
for all functions $g,h$.
Therefore, if we want to express independence statements we need to consider a space of functions which includes all products $gh$. Since $L$ must be defined on a real vector space of functions, a `minimal' way to do that is to consider:
\begin{equation}
 \label{eq:gambG0}
\mathcal{G}:=\text{span}\{ g(\x,\x^\dagger)h(\y,\y^\dagger): \text{ for all } f \in \mathcal{F}, h \in \mathcal{H}\}.
\end{equation}
Notice that $\mathcal{G}$ is a vector space that contains the constants. Moreover, we have that  $\mathcal{F},\mathcal{H} \subset \mathcal{G}$. In fact, for $H=I$, one has ${\bf x}^\dagger G {\bf x}{\bf y}^\dagger H {\bf y}={\bf x}^\dagger G {\bf x}$ and vice versa. In this `product space', independence judgements can be expressed in terms of expectations by stating $L({\bf x}^\dagger G {\bf x}{\bf y}^\dagger H {\bf y})=L({\bf x}^\dagger G {\bf x})L({\bf y}^\dagger H {\bf y})$. 

The following proposition shows how the tensor-product arises in QT.

\begin{proposition}
\label{prop:locald}
 Equation~\eqref{eq:gambG0} can 
 be rewritten as
\begin{equation}
 \label{eq:gambG}
\mathcal{G}=\{({\bf x}\otimes{\bf y})^\dagger G ({\bf x}\otimes {\bf y}) : G \text{ is a $n_xn_y\times n_xn_y$ Hermitian}\},
\end{equation}
where $\otimes$ is the Kronecker product.
\end{proposition}

We can then prove the following result, that can be straightforwardly extended to any multipartite system.

\begin{theorem}[Representation theorem for two particle systems]
\label{th:reprdouble}
For every   $g([{\bf x},\y],[{\bf x}^\dagger,\y^\dagger])=({\bf x} \otimes {\bf y})^\dagger G({\bf x} \otimes {\bf y}) \in \mathcal{G}$,
%in \eqref{eq:gambG}, 
the following definitions are equivalent
\begin{enumerate}
\item $\widetilde{L}:\mathcal{G}\rightarrow \reals$ is a QEO 
with $\mathcal{C}^+=\Sigma^+_{n_x n_y}$, where 
\begin{equation}
\label{eq:SOScone}
\Sigma^+_{n_x n_y}=\{({\bf x} \otimes {\bf y})^\dagger H({\bf x} \otimes {\bf y}): ~~H\succeq 0\},
\end{equation}
is the so-called closed-convex cone of Sum-of-Squares (SOS) Hermitian polynomials (of dimension $n_x n_y$).
\item $\widetilde{L}$ can be written as 
\begin{equation}
 \label{eq:expM}
\widetilde{L}(g)= Tr(GM),
\end{equation}
where $M$ is a $n_x \times n_x$ Hermitian matrix such that $M\succeq 0, ~~Tr(M)=1$.
\end{enumerate}
Furthermore, $\widetilde{L}$ is a tractable QEO.
\end{theorem}
% This result can be straightforwardly extended to any multipartite system. 
Let $\underline{\lambda}_G$, resp. $\overline{\lambda}_G$, be the minimal, resp. the maximal, eigenvalue of $G$.
It can be verified that the definition provided by Equation~\eqref{eq:expM} in the theorem above
%the above definition of QEO 
implies that
$\widetilde{L}(g)\geq \underline{c}_g$, with $\underline{c}_g:=\underline{\lambda}_G$.
Moreover, from property \eqref{eq:infsupboundgen1}, we can  derive
\begin{equation}
\label{eq:infsupboundQM1}
\underline{\lambda}_G \leq \inf g  \leq \sup g \leq \overline{\lambda}_G.
\end{equation}
%where $\overline{\lambda}_G$ is the maximum eigenvalue of $G$.
Since these inequalities can be strict, %this shows that 
density matrices %$\rho=M$ 
are \textbf{quasi-moment matrices}, where `quasi' means that their underlying linear operator is a QEO.

 \begin{example}
 \label{ex:entQT}
 Consider the case $n_x=n_y=2$, and the matrix
 \begin{equation}
\label{eq:exlinearoperator}
\begin{aligned}
&\widetilde{L}\left((\x \otimes \y)(\x \otimes \y)^\dagger\right)=\\ &\widetilde{L}\left(\left[\begin{smallmatrix}x_{1} x_{1}^{\dagger} y_{1} y_{1}^{\dagger} & x_{1}^{\dagger} x_{2} y_{1} y_{1}^{\dagger} & x_{1} x_{1}^{\dagger} y_{1}^{\dagger} y_{2} & x_{1}^{\dagger} x_{2} y_{1}^{\dagger} y_{2}\\x_{1} x_{2}^{\dagger} y_{1} y_{1}^{\dagger} & x_{2} x_{2}^{\dagger} y_{1} y_{1}^{\dagger} & x_{1} x_{2}^{\dagger} y_{1}^{\dagger} y_{2} & x_{2} x_{2}^{\dagger} y_{1}^{\dagger} y_{2}\\x_{1} x_{1}^{\dagger} y_{1} y_{2}^{\dagger} & x_{1}^{\dagger} x_{2} y_{1} y_{2}^{\dagger} & x_{1} x_{1}^{\dagger} y_{2} y_{2}^{\dagger} & x_{1}^{\dagger} x_{2} y_{2} y_{2}^{\dagger}\\x_{1} x_{2}^{\dagger} y_{1} y_{2}^{\dagger} & x_{2} x_{2}^{\dagger} y_{1} y_{2}^{\dagger} & x_{1} x_{2}^{\dagger} y_{2} y_{2}^{\dagger} & x_{2} x_{2}^{\dagger} y_{2} y_{2}^{\dagger}\end{smallmatrix}\right]\right)=\frac{1}{2}\begin{bmatrix}
      0 & 0 & 0 &0\\
      0 & 1 & 1 &0\\
      0 & 1 & 1 &0\\
      0 & 0 & 0 &0\\
     \end{bmatrix}.
\end{aligned}
\end{equation}
We want to show that the above density matrix is a quasi-moment matrix and not a classical moment-matrix. Since the matrix has rank one, in order to write  $\widetilde{L}\left((\x \otimes \y)(\x \otimes \y)^\dagger\right)$ as a classical expectation, we need to find an atomic probability measure (a Dirac's delta) 
$\delta_{\tilde{\x}}(\x)\delta_{\tilde{\y}}(\y)$ for some $\tilde{\x},\tilde{\y}$, such that
$$
\int (\x \otimes \y)(\x \otimes \y)^\dagger \delta_{\tilde{\x}}(\x)\delta_{\tilde{\y}}(\y)=\frac{1}{2}\begin{bmatrix}
      0 & 0 & 0 &0\\
      0 & 1 & 1 &0\\
      0 & 1 & 1 &0\\
      0 & 0 & 0 &0\\
     \end{bmatrix}.
$$
Then we must choose $\tilde{\x},\tilde{\y}$ so that
$$
\tilde{\x} \otimes \tilde{\y}=\begin{bmatrix}
\tilde{x}_1\tilde{y}_1\\
\tilde{x}_1\tilde{y}_2\\
\tilde{x}_1\tilde{y}_1\\
\tilde{x}_2\tilde{y}_2\\
\end{bmatrix}=\begin{bmatrix}
0\\
*\\
*\\
0\\
\end{bmatrix}
$$
for some $*$ different from zero. This is impossible, because the first row implies that either $\tilde{x}_1=0$ or $\tilde{y}_1=0$ and the last row that either $\tilde{x}_2=0$ or $\tilde{y}_2=0$. Taken together, these constraints imply that $*=0$, meaning that the  QEO $\widetilde{L}$ cannot be written as a classical expectation operator.
 \end{example}
The previous example enables us to understand why in general $\widetilde{L}(g)$ is not a classical expectation.  In fact, given $g([{\bf x},{\bf y}],[{\bf x}^\dagger,{\bf y}^\dagger]):=({\bf x}\otimes{\bf y})^\dagger G ({\bf x}\otimes {\bf y})$, we can find  a  PSD Hermitian matrix $M$ of trace one such that 
 $$
 Tr(GM)< \inf g.
 $$
 Hence, by taking $\widetilde{L}(\cdot) :=Tr(\cdot M)$, the linear operator satisfies Equation~\eqref{eq:expM0} in Theorem~\ref{th:reprsingle} but, since it does not satisfies property \eqref{eq:A},  it is not is a valid expectation operator.
% The matrix $M$ in Theorem~\ref{th:reprdouble} therefore violates property \eqref{eq:A}, meaning that in this case $L(\cdot) =Tr(\cdot M)$ is not a valid expectation operator for every $M\succeq 0$ with $Tr(M)=1$.
 To to satisfy this property, some addition constraints for $M$ must hold.
 %satisfy some additional constraints besides being PSD and trace one .
 Determining these constraints is an \textbf{NP}-hard problem, and actually entails to prove that $M$ is separable.
 This is due to the fact that, when considering a multipartite system, classical expectation operators are not tractable.
 
\begin{proposition}[\cite{gurvits2003classical}]
Computing the minimum of a function belonging to $\mathcal{G}$
%as described by Equation~\eqref{eq:gambG} 
is \textbf{NP}-hard. %Therefore, $L$
\end{proposition}

Instead,  QT is a tractable QEO.  
Indeed, the corresponding claim stated in Theorem~\ref{th:reprdouble} follows immediately from the fact that the membership problem $g-c \in \Sigma^+_2$ can be solved using semi-definite programming. % (SDP). 
%and this explains the tractability.

\subsection{Representation theorem for probability}
As we will discuss in Section \ref{sec:defin}, a representation theorem {\`a} la de Finetti expresses a classical probability distribution  in terms of a (quasi-)expectation of certain polynomial functions.

We can obtain a similar result in QT using the setting underlying Gleason's theorem. For simplicity, let us focus again only on a generic system composed by two particles denoted respectively by $\x \in \overline{\mathbb{C}}^{n_x}$ and by $\y \in \overline{\mathbb{C}}^{n_y}$.
Let $\mathfrak{P}(\overline{\mathbb{C}}^{n_z})$ be the lattice of orthogonal projectors on $\overline{\mathbb{C}}^{n_x}$ with $n_z=n_xn_y$.
In QT, a valid probability measure $P: \mathfrak{P}(\overline{\mathbb{C}}^{n_z}) \to [0,1]$  has to satisfy the following constraints/symmetries:
\begin{align}
\label{eq:p1gl}
&P\left(\mathbf{z}_1 \vee \mathbf{z}_2 \vee \dots \vee \mathbf{z}_{n_z}\right)=1, \tag{P1}\\
\label{eq:p2gl}
&P\left(\mathbf{z}_1 \vee \mathbf{z}_2 \vee \dots \vee \mathbf{z}_{m}\right)=\sum^{m}_{i=1} P(\mathbf{z}_i), \tag{P2}
\end{align}  
for each sequence $(\mathbf{z}_1, \dots, \mathbf{z}_{m})$ of mutually orthogonal directions, and $m\leq n_z$.

The next results, which is an immediate consequence of Theorem~\ref{th:reprdouble}, tell us, again, that QT can be expressed as a QEO.

\begin{corollary}[Representation theorem for probabilities over the lattice of orthogonal projectors]
\label{co:gleason}
Let  $P$ be a probability satisfying \eqref{eq:p1gl}--\eqref{eq:p2gl}. There exists a QEO $\widetilde{L}$ as in Theorem~\ref{th:reprdouble} such that:
\begin{equation}
\label{eq:probGleason}
P(\mathbf{z}_i)=\widetilde{L}\left(({\bf x}\otimes{\bf y})^\dagger \mathbf{z}_i\mathbf{z}^\dagger_i ({\bf x}\otimes {\bf y})\right).
\end{equation}
\end{corollary}

%\begin{remark}   
The previous Corollary~\ref{co:gleason} underlines that, similar to GPTs, our approach aims at relating the observed probabilities to  some belief states. We however differ from GPTs in the way these probabilities are represented. We see probabilities as quasi-expectations of polynomials:
$$
\textit{probability}= \widetilde{L}(\textit{polynomials}).
$$
The functional $\widetilde{L}: \mathcal{F}\rightarrow \reals$ is a real-valued operator defined on the space of real-valued functions $\mathcal{F}$. Therefore, as in GPTs, QEOs are defined on a real-vector space. This means that, despite the aforementioned difference, the two approaches are formally equivalent when expressed using order unit spaces and using duality \cite[Th.3]{Benavoli2021f}.

In general, probabilities can always be expressed as expectation of indicator functions, $P(\x \in A)=E[I_{A}(\x)]$, where $I_{A}(\x)=1$ if $\x \in A$ and zero otherwise. Instead, \eqref{eq:probGleason} relates the probabilities of the outcome of a quantum experiment to the expectation of a polynomial function of certain `hidden-variables' $\x,\y$.
As we will explain in the next section, the expression \eqref{eq:probGleason}  is similar to the way we model the rolls of a dice. For instance, the probability of the result (face1, face2) of two rolls of a dice can be expressed  as an expectation of a polynomial $P(\text{face}~1, \text{face}~2)=L(\theta_1 \theta_2)=\int \theta_1 \theta_2 dp(\theta_1,\dots,\theta_6)$, where  $\theta_i$ is the probability of face $i$ (hidden-variables). $p(\theta_1,\dots,\theta_6)$ expresses our belief about the bias of the dice (equivalently, a preparation-procedure). The difference is that $\widetilde{L}$ in  \eqref{eq:probGleason}  is a quasi-expectation operator but, as we will explain in this section, quasi-expectation operators appear also in the representation theorem for the
probability of finitely exchangeable rolls of a classical dice.

%Finally, notice that our approach is a representation not a reconstruction.  This notwithstanding, it may be easily turned in a quantum reconstruction. For instance, Proposition~\ref{prop:locald} can be shown to be a consequence of  tomographic locality or local discriminability, see for instance \cite{dariano_chiribella_perinotti_2017}. But this is out of the scope of the present work.
%\end{remark}

\section{Exchangeability}
In the previous section, we provided two representation results (Theorem~\ref{th:reprdouble} and Corollary~\ref{co:gleason}) for distinguishable particles via QEOs. In the second part of this work we derive the \textit{first and second quantisation} in QT for bosons  through the statistical concept of \textit{exchangeable sequence of random variables}.

Despite being usually considered a truly `weird' quantum phenomenon, particles indistinguishability is actually not so special. As a matter of fact, in Section \ref{sec:defin} we show that exchangeable probabilities for a classical dice are as weird as QT. Starting from de Finetti's representation theorem, we discuss a series of representation theorems for the probability of finitely exchangeable rolls of a classical dice. Analogously to what happens in QT, these representation theorems involve negative probabilities, or equivalently, QEOs. We then discuss how the `weirdness' disappears as the considered number of dice rolls increases.
Finally, we use the same approach to derive a representation theorem for the second quantisation for bosons. In doing so, we show how classical reality emerges in QT as the considered number of identical bosons increases.

\subsection{Exchangeability for classical dices is not so classical}
\label{sec:defin}
We  present de Finetti's approach to exchangeability with an example. Consider a dice whose possibility space is $\Omega=\{d_1,d_2,d_3,d_4,d_5,d_6\}$ (the six faces of the dice) and denotes with $t_1,t_2,\dots,t_r$ the results of $r$-rolls of the dice.

\begin{definition}
\label{def:excg}
A sequence of variables $t_1,t_2,\dots,t_r$ is said to be finitely exchangeable, if their joint probability satisfies
$$
P(t_1,t_2,\dots,t_r)=P(t_{\pi_1},t_{\pi_2},\dots,t_{\pi_r}),
$$
for any permutation $\pi$ of the indexes.
\end{definition}
%Equivalently, we can formulate it in terms of expectations. Consider for simplicity $r=2$ and denotes with $\theta_{ij}=p(t_1=i,t_2=j)$, that is the probability of the outcome  face-$i$ in the first roll and face-$j$ in the second roll. Then, $t_1,t_2$ is said to be exchangeable whether
%$$
%L(\theta_{ij})=L(\theta_{ji}),
%$$
%for all $i,j=1,2,\dots,6$.

This definition of exchangeability expressed in terms of symmetry to label-permutation is formally equivalent to the first quantisation in QT. De Finetti also introduced the second quantisation. Given $t_1,t_2,\dots,t_r$ are exchangeable, the output of the $r$-rolls is fully characterised by the counts:
$$
\underset{n_{1}}{d_1},~\underset{n_{2}}{d_2},~\dots~,\underset{n_{6}}{d_6},
$$
where $n_i$ denotes the number of times the dice landed on face $d_i$ in the $r$-rolls. We can represent the counts as a vector  $[n_1,n_2,\dots,n_6]$.
De Finetti then proved his famous\footnote{De Finetti proved his theorem for the binary case, a coin, but this result can easily be extended to the dice.} \textit{Representation Theorem}.
\begin{proposition}[\cite{finetti1937}]
\label{prop:definetti}
If $t_1,t_2,t_3,\dots$ is an infinitely exchangeable sequence of random variables (that is a sequence that satisfies Definition \ref{def:excg} for every $r$) defined in the possibility $\{d_1,d_2,\dots,d_6\}$ and
which has probability  measure $P$, then there exists a distribution function $q$ such that
\begin{equation}
    \label{eq:reprTh0}
P(t_1,\dots,t_n) = \int_\Theta \theta_1^{n_1}\theta_2^{n_2}\cdots \theta_5^{n_5}\left(1-\theta_1-\dots-\theta_6\right)^{n_6} dq({\boldsymbol \theta}), 
\end{equation}
where ${\boldsymbol \theta}^\top=[\theta_1,\theta_2,\dots,\theta_5]$ are the probabilities of the corresponding faces, $\Theta$ is the possibility space for ${\boldsymbol \theta}$, and $n_i$ is the number of times the dice landed on the i-th face in the n rolls.
\end{proposition}
This theorem is usually interpreted as stating that a sequence of random variables is exchangeable if it is conditionally independent and identically distributed. For a fixed ${\boldsymbol \theta}$, this for instance means that 
$P(t_1=2,t_2=3,t_3=2,t_4=1)=\theta_1\theta_2^2\theta_3$ (the product comes from the independence assumption). For an unknown ${\boldsymbol \theta}$, $P$ is an infinite mixture of $\theta_1\theta_2^2\theta_3$, which depends on our beliefs over ${\boldsymbol \theta}$ expressed by $q({\boldsymbol \theta})$.

The polynomials
\begin{equation}
    \label{eq:Berne}
\left\{\theta_1^{n_1}\theta_2^{n_2}\cdots \theta_5^{n_5}\left(1-\theta_1-\dots-\theta_6\right)^{n_6}: \sum_{i=1}^6 n_i=n\right\},
\end{equation}
where $n$ is the number of rolls,  are called \textit{multivariate Bernstein polynomials} and play a central role in proving de Finetti's Representation Theorem. 
They satisfy a set of useful properties:
\begin{itemize}
    \item  Bernstein  polynomials of fixed degree $n$ form a
basis for the linear space of all polynomials whose degree is at most $n$;
\item Bernstein  polynomials form a partition of unity:
\begin{equation}
    \label{eq:Bernsum1}
\sum_{\substack{[n_1,\dots,n_6]:\\\sum_{i=1}^6 n_i=n}} \theta_1^{n_1}\theta_2^{n_2}\theta_3^{n_3}\theta_4^{n_4}\theta_5^{n_5} \left(1-\theta_1-\dots-\theta_6\right)^{n_6}=1.
\end{equation}
for every $n$.
\end{itemize}

Reasoning about exchangeable variables $t_i$ can be reduced to reasoning
about count vectors or polynomials of frequency vectors (that is, Bernstein polynomials) \cite{kerns2006definetti,harper2007probability,de2012exchangeability}. Working with this polynomial representation automatically guarantees
that exchangeability is satisfied, without having to go back to the more complex world of labeled variables $t_i$.
 
 It is well known that de Finetti's theorem  does not hold in general for finite sequences of exchangeable random variables. In such case, one can only prove the following representation theorem.
 \begin{proposition}[\cite{kerns2006definetti}]
\label{prop:kerns}
Given a finite sequence of exchangeable variables $t_1,t_2,\dots,t_r$, there exists a signed measure  $\nu$, satisfying  $\nu(\Theta)=1$, such that:
 \begin{equation}
    \label{eq:reprTh}
P(t_1,\dots,t_r) = \int_\Theta \theta_1^{n_1}\theta_2^{n_2}\cdots \theta_5^{n_5}\left(1-\theta_1-\dots-\theta_6\right)^{n_6} d\nu({\boldsymbol \theta}).
\end{equation}
\end{proposition}
Signed measure  means that $\nu$  includes some negative probabilities. Note that, $P$ is always a valid probability mass function.
%In other words, 
% \begin{equation}
%    \label{eq:Ltildedefin}
%\widetilde{L}(\theta_1^{n_1}\theta_2^{n_2}\cdots %\theta_6^{n_6}):=
%% \int_\Theta 
 %\theta_1^{n_1}\theta_2^{n_2}\cdots %\theta_6^{n_6}d\nu({\boldsymbol \theta})
 %\end{equation}
 %is not a classical expectation operator. 
 To see that, consider the case $r=2$ and
 \begin{equation}
     \label{eq:psym1}
      P(t_1=i,t_2=j)= P(t_1=j,t_2=i)=\frac{1}{30},
 \end{equation}
 for all $i \neq j =1,2,\dots,6$.  This implies that
  \begin{equation}
     \label{eq:psym2}
P(t_1=i,t_2=i)=0
\end{equation}
 for all $i$. The   constraints \eqref{eq:psym2} cannot be satisfied by a classical expectation operator. In fact,  $0=P(t_1=i,t_2=i)=\int_\Theta \theta_1^{n_1}\theta_2^{n_2}\cdots \left(1-\theta_1-\dots-\theta_6\right)^{n_6} dq({\boldsymbol \theta}) $ for all $i$ would imply that $q$ puts mass $1$ at all the $\theta_i=0$, which is impossible. This means that, although the $P$ in \eqref{eq:psym1} is a valid probability, we cannot find any hidden-variable theory (any $q({\boldsymbol \theta})$) which is compatible with it. This is similar to what happens in QT with entanglement, as discussed in Example \ref{ex:entQT}. 
 
  Note that, in  Equation~\eqref{eq:reprTh}, the only valid  signed-measures $\nu$ are those which give rise to exchangeable classical probabilities $P(t_1,t_2,\dots,t_r)$.  
These valid signed-measures can be found by solving a linear programming problem. This fact is a consequence of the following \textit{representation theorem}.
  \begin{proposition}[ \cite{de2012exchangeability}]
  \label{prop:LPL}
  Consider the vector space of functions $$\mathcal{F}=\text{span}(\{\theta_1^{n_1}\theta_2^{n_2}\theta_3^{n_3}\theta_4^{n_4}\theta_5^{n_5} \left(1-\theta_1-\dots-\theta_6\right)^{n_6}: \sum_{i=1}^6 n_i=r\})$$ and a QEO $\widetilde{L}: \mathcal{F} \rightarrow \mathbb{R}$ defined by: 
       \begin{equation}
     \label{eq:LPcone0}
     \begin{aligned}
          \underline{c}_g=\arg \sup c ~\text{ s.t. }~ g-c \in \mathcal{B}^+_r,
     \end{aligned}
     \end{equation}
for all $g \in \mathcal{F} $, where $c$ denotes the constant function  and
\begin{equation}
 \label{eq:LPcone1}
\begin{aligned}
    \mathcal{B}^+_r =\Bigg\{ & \sum\limits_{{\bf n}= [n_1,\dots,n_6]: \sum_i n_i=r} \hspace{-3mm}u_{{\bf n}} 
  \theta_{1}^{n_1} \theta_{2}^{n_2}\cdots  \left(1-\theta_1-\dots-\theta_6\right)^{n_{6}}:\\& u_{{\bf n}}\in \mathbb{R}^{+}  \Bigg\}.
\end{aligned}
\end{equation}
Then any exchangeable $P(t_1,t_2,\dots,t_r)$ can be written as $\widetilde{L}(\theta_1^{n_1}\theta_2^{n_2}\cdots \theta_6^{n_6})$.
 \end{proposition}
 Note that, thanks to the partition-of-unit property of Bernstein's polynomials, the vector space $\mathcal{F}$ includes the constants :
 $$
 c\sum_{\substack{[n_1,\dots,n_6]:\\\sum_{i=1}^6 n_i=r}} \theta_1^{n_1}\theta_2^{n_2}\theta_3^{n_3}\theta_4^{n_4}\theta_5^{n_5} \left(1-\theta_1-\dots-\theta_6\right)^{n_6}=c.
 $$
 
 Proposition~\ref{prop:LPL} states  that the valid $\nu$s are only the ones for which $\int_\Theta g({\boldsymbol \theta}) d\nu({\boldsymbol \theta})$ satisfies  (A$^*$) with $\underline{c}_g$ defined as in Equation  \eqref{eq:LPcone0}. The cone
 $\mathcal{C}^+_r$ is called the closed-convex cone of \textit{nonnegative Bernstein polynomials} of degree $r$. For this cone, the optimisation problem stated in Equation~\eqref{eq:LPcone0} can be solved in \textbf{P}-time (by linear programming). Indeed,  the membership $g-c \in \mathcal{C}^+_r$ can be verified by  checking that the expansion of the polynomial $g-c$ with respect to the Bernstein basis has all nonnegative coefficients. Therefore,  $\widetilde{L}$ is a tractable QEO.
 
 \begin{remark}
 Formally, Proposition~\ref{prop:LPL} is the equivalent to Gleason's theorem (Corollary~\ref{co:gleason}) for finitely exchangeable rolls of a classical dice. By comparing Theorem~\ref{th:reprdouble} and Equation~\eqref{eq:SOScone} with Proposition~\ref{prop:LPL} and Equation~\eqref{eq:LPcone1}, the reader can understand the differences and similarities between these two representation theorems.
 \end{remark}

In \cite{zaffalon2019b}, we have provided a `theory of probability' built upon the QEO \eqref{eq:LPcone0} displaying the same weirdness attributed to QT. The following is a similar example.

\begin{example}
\label{eq:LP}
Assume we roll the dice twice and consider the following strictly nonnegative polynomial of ${\boldsymbol \theta}$:
$$
g({\boldsymbol \theta})=\theta_1^2-\theta_1\theta_2+\theta_2^2+0.05>0,
$$
Note that, $\min_{{\boldsymbol \theta} \in \Theta} g =0.05$. Observe also that  the coefficients of the expansion of the polynomial $g({\boldsymbol \theta})$ are not all nonnegative and, therefore, $g$ does not belong to $\mathcal{B}^+_2$. We can then show that $g$ is an \textbf{entanglement witness} for  the QEO defined in Proposition~\ref{prop:LPL}, that is there exists an  $\widetilde{L}$  such that $\widetilde{L}(g)<0$.
We can find the worst $\widetilde{L}$, that is the `maximum entangled' QEO,  by solving the following optimisation problem
\begin{equation}
\label{eq:optimdice}
\begin{aligned}
\underline{c}_g=&\max\, c
~~~s.t.~~~g-c \in \mathcal{B}^+_2
\end{aligned}
\end{equation}
where
\begin{equation}
\label{eq:coneposex}
\begin{aligned}
&\mathcal{B}^+_2=\{u_{200000}\theta_1^2 +  u_{110000}\theta_1\theta_2+  u_{101000}\theta_1\theta_3 + \\
& u_{100100}\theta_1\theta_4 +  u_{100010}\theta_1\theta_5+  u_{100001}\theta_1(1-\theta_1-\dots-\theta_5) + \\
&u_{020000}\theta_2^2 +  \dots+ u_{000011}\theta_5(1-\theta_1-\dots-\theta_5)\\
&+ u_{000002}(1-\theta_1-\dots-\theta_5)^2: u_{[n_1,n_2,\dots,n_6]}\geq0\}.
\end{aligned}
\end{equation}
 The solution of \eqref{eq:optimdice} can be computed
 by solving the following linear programming problem:
\begin{equation}
\begin{aligned}
\max_{c\in \reals,u_{[n_1,n_2,\dots,n_6]}\in \reals^+} c & \\
 -u_{000002} + u_{100001} - u_{200000} + 1 =0\\
 -2u_{000002} + u_{010001} + u_{100001} - u_{110000} - 1=0\\
 2u_{000002} - u_{100001}=0\\
 -u_{000002} + u_{010001} - u_{020000} + 1=0\\
 2u_{000002} - u_{010001}=0\\
 - c - u_{000002}+0.05=0
\end{aligned}
\end{equation}
where the equality constraints have been obtained by equating the coefficients of the monomials
in 
$$
\begin{aligned}
&g(\theta)-c =u_{200000}\theta_1^2 +  u_{110000}\theta_1\theta_2+  u_{101000}\theta_1\theta_3 + \\
& u_{100100}\theta_1\theta_4 +  u_{100010}\theta_1\theta_5+  u_{100001}\theta_1(1-\theta_1-\dots-\theta_5) + \\
&u_{020000}\theta_2^2 +  \dots+ u_{000011}\theta_5(1-\theta_1-\dots-\theta_5)\\
&+ u_{000002}(1-\theta_1-\dots-\theta_5)^2.
\end{aligned}
$$
For instance, consider the the constant term in the r.h.s.\ of the above equation: that is $u_{000002}$. It must be equal to the constant term of $g(\theta)-c$, that is $0.05-c$. Therefore, we have that
$$
0.05-c-u_{000002}=0.
$$
Similarly, consider the coefficient of the monomial $\theta_1^2$: this is $u_{200000}-u_{100001}+u_{200002}$. It must be equal to the coefficient of the monomial $\theta_1^2$ in $g(\theta)-c$, which is $1$, that is
$$
1-(u_{200000}-u_{100001}+u_{200002})=0.
$$
The other constraints can be obtained in a similar way.

The solution of \eqref{eq:coneposex} is $$
\underline{c}_g=-0.45.
$$
The QEO which attains the above solution can be found via duality:
$$
\begin{aligned}
\widetilde{L}(\theta_1^2)=0.125,~~
\widetilde{L}(\theta_1\theta_2)=0.75,~~
\widetilde{L}(\theta_2^2)=0.125,
\end{aligned}
$$
and zero otherwise. Given
$$
\begin{aligned}
&\widetilde{L}(\theta_1^2-\theta_1\theta_2+\theta_2^2+0.05)\\
&=\widetilde{L}(\theta_1^2)-\widetilde{L}(\theta_1\theta_2)+\widetilde{L}(\theta_2^2)+0.05=-0.45<0,
\end{aligned}
$$
This proves that $\widetilde{L}$ is `entangled'.
\end{example}
Given a finitely exchangeable sequence $t_1,t_2,\dots,t_r$ and the corresponding exchangeable probability $p$, assume we  focus on the results of any two rolls, generically denoted as $t_a,t_b$.
Equivalently, given the probability
$$
P(t_1,t_2,\dots,t_r)=P(t_{\pi_1},t_{\pi_2},\dots,t_{\pi_r}),
$$
we are interested on the marginal
$$
P(t_a,t_b)=P(t_{\pi_a},t_{\pi_b}).
$$
It does not matter what $t_a,t_b$ are because all the marginals
$P(t_1,t_2)=P(t_1,t_3)=\dots$ are identical. We denote the system composed by $t_1,t_2,\dots,t_r$  as $\mathcal{A}_r$ and the marginal system $t_a,t_b$ as $\mathcal{A}_{r|2}$. 

We can then prove the following theorem for exchangeable dices.
\begin{theorem}
\label{th:classidiceconvergebnce}
$\mathcal{A}_{r+1|2}$ is more classical then $\mathcal{A}_{r|2}$ for any $r$. $\mathcal{A}_{r|2}$ becomes classical when $r\rightarrow \infty$.
\end{theorem}
Here, `more classical' means that for any witness $g$ of degree $2$ we have that $\inf_{\widetilde{L}} \widetilde{L}_{\mathcal{A}_{r+1|2}}(g)>\inf_{\widetilde{L}} \widetilde{L}_{\mathcal{A}_{r|2}}(g)$. Moreover, $\widetilde{L}$ becomes a classical expectation operator for $r\rightarrow \infty$.

To prove and clarify this theorem, we introduce the concept of \textit{degree extension}. In the previous example, we considered two rolls of a dice and focused on the polynomial $\theta_1^2-\theta_1\theta_2+\theta_2^2+0.05$ for inference. However, we can equivalently  consider an experiment involving $r>2$ rolls, but keeping the focus on the same inference. This is possible because for any polynomial $g$:
{\small
\begin{equation}
\label{eq:LPextension}
\begin{aligned}
g({\boldsymbol \theta})&=g({\boldsymbol \theta})\left(\sum_{\substack{[n_1,\dots,n_6]:\\ \sum_{i=1}^6 n_i=r}} \theta_1^{n_1}\theta_2^{n_2}\theta_3^{n_3}\theta_4^{n_4}\theta_5^{n_5} \left(1-\theta_1-\dots-\theta_6\right)^{n_6}\right),
\end{aligned}
\end{equation}}
which holds because the term between bracket is equal to one.
By doing that,  we can embed any polynomial $g$ of degree $s$ in the space of polynomials of degree $s+r$. 

\begin{lemma}
\label{lem:lemwitndice}
Consider a $s$-degree polynomial $g({\boldsymbol \theta})$. The following conditions are equivalent.
\begin{enumerate}
    \item $g({\boldsymbol \theta})>0$ for all   ${\boldsymbol \theta} \in \Theta$;
    \item there exist positive integer $r$ such that the polynomial \\
      $g({\boldsymbol \theta})\left(\sum\limits_{\substack{[n_1,\dots,n_6]:\\ \sum_{i=1}^6 n_i=r}} \theta_1^{n_1}\theta_2^{n_2}\theta_3^{n_3}\theta_4^{n_4}\theta_5^{n_5} \left(1-\theta_1-\dots-\theta_6\right)^{n_6}\right)$     is in $\mathcal{B}^+_{r+s}$.
\end{enumerate}
\end{lemma}
Consider the case $s=2$ as in Example \ref{eq:LP}. Lemma~\ref{lem:lemwitndice} tells us that for any witness $g$ such that  ${L}(g)>0$, there exists $r$ such that $\widetilde{L}_{\mathcal{A}_{r|2}}(g)>0$.

\begin{figure}[!htp]
\centering
\includegraphics[width=8cm]{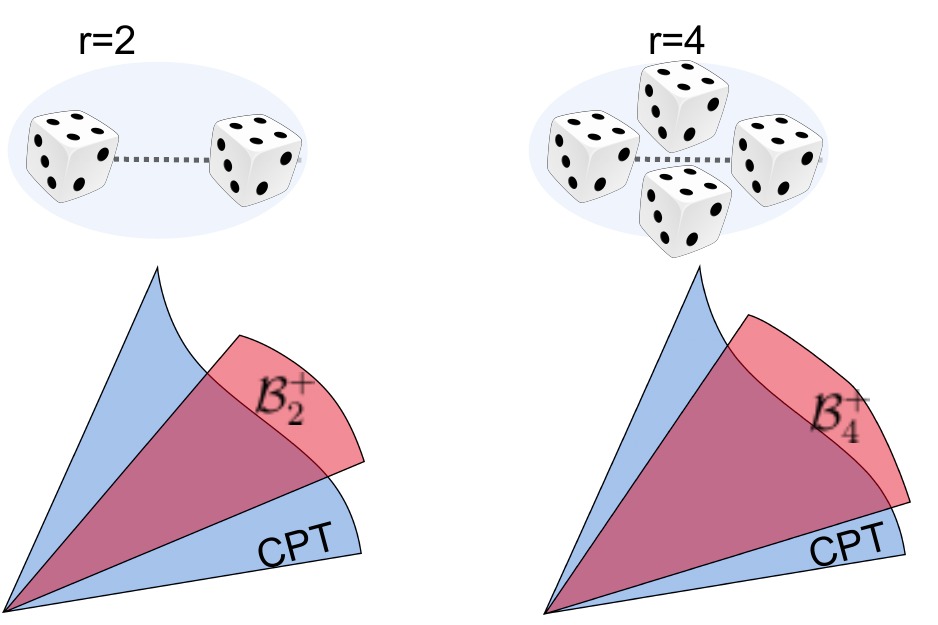}
\caption{Pictorial representation of the convergence of the Bernstein cone of nonnegative polynomials to the cone of nonnegative polynomials (CPT)  for $r$ exchangeable dices.}
\label{fig:conedice}
\end{figure}

\begin{example}
Let us go back to Example \ref{eq:LP}. Figure \ref{fig:LPcon} reports the minimum of $\widetilde{L}_{\mathcal{A}_{r|2}}(g)=\widetilde{L}_{\mathcal{A}_{r|2}}(\theta_1^2-\theta_1\theta_2+\theta_2^2+0.05)$ as a function of $r\geq2$.
 This shows that $\widetilde{L}_{\mathcal{A}_{r|2}}(g)$ quickly tends  to the classical limit $0.05$ at the increase of the degree $r$.
  \begin{figure}[htp!]
 \centering
  \includegraphics[width=7cm]{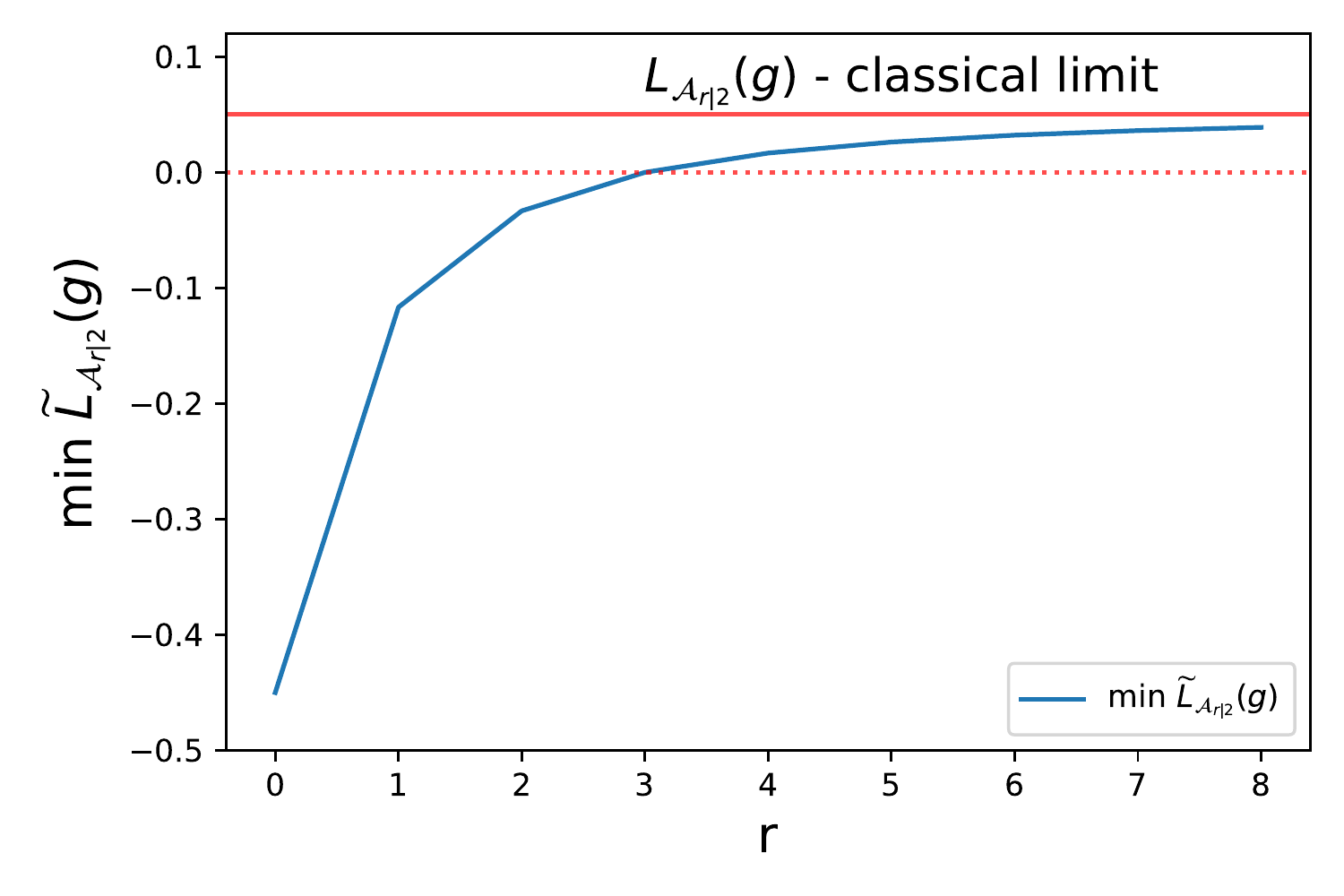}
  \caption{Classical dice: convergence of $\widetilde{L}_{\mathcal{A}_{r|2}}(g)$  to ${L}(g)$ at the increase of  $r$.}
  \label{fig:LPcon}
 \end{figure}
 In other words, the system behaves more classically at the increase of $r$. This is due to the fact the cone  $ \mathcal{B}^+_r$ is included in the cone $\mathcal{B}^+_{r+1}$, as pictorially depicted in Figure \ref{fig:conedice}, and converges to the cone of nonnegative polynomials of the variable ${\boldsymbol \theta}$ as  $r \rightarrow \infty$.
\end{example}

The moral exemplified by the previous example is that for large $r$ the exchangeable probability $P$ becomes compatible with a hidden-variable theory. We will see in the next section that the same type of convergence is responsible of the emergence of classical reality in QT.

\subsection{Exchangeability of identical particles} 
Having connected density matrices with QEOs, we can easily derive the symmetrisation  postulate for identical particles \cite{benavoli2021indi} by imposing exchangeability constraints on the operator, analogously to what done in CPT.
%Given exchangeability is a linear  constraint and $\widetilde{L}$ is linear, to deal with identical particles, we can use the formalism derived by de Finetti in CPT in his famous \textit{representation theorem}   \cite{finetti1974-5}.
For instance, for two identical particles, an exchangeable QEO is defined as follows.
\begin{definition}
Let $\widetilde{L}:\mathcal{G}\rightarrow \mathbb{R}$ be a linear operator  satisfying property \eqref{eq:Astar}. If, for each polynomial $({\bf x}_1 \otimes {\bf x}_2)^\dagger G({\bf x}_1 \otimes {\bf x}_2)$, $\widetilde{L}$ satisfies the constraints
{\small
\begin{align}
\label{eq:indicon1}
 &\widetilde{L}(({\bf x}_1 \otimes {\bf x}_2)^\dagger G({\bf x}_1 \otimes {\bf x}_2))=\widetilde{L}(({\bf x}_2 \otimes {\bf x}_1)^\dagger G({\bf x}_2 \otimes {\bf x}_1))~\\
 \label{eq:indicon2}
 &=\delta_{*}\widetilde{L}\left(\tfrac{1}{2}[({\bf x}_2 \otimes {\bf x}_1)^\dagger G({\bf x}_1 \otimes {\bf x}_2)+({\bf x}_1 \otimes {\bf x}_2)^\dagger G({\bf x}_2 \otimes {\bf x}_1)]\right)
\end{align}}
where $\delta_{*}$ is the sign of the permutation, then $\widetilde{L}$ is called symmetric  if $\delta_{*}=1$  (bosons) or anti-symmetric if $\delta_{*}=-1$   (fermions).
\end{definition}
By linearity, these equalities can be translated into constraints on the valid density matrices (previously denoted as $M$) under exchangeability \cite{benavoli2021indi}:
$$
\rho= \Pi_{\star} \rho \Pi_{\star},
$$
where $\Pi_{\star}$ is the symmetriser ($\star=Sym$) for bosons and anti-symmetriser  ($\star=Anti$) for fermions. The result follows by first  assuming that  ${\bf x}_1 , {\bf x}_2$ are exchangeable \cite{benavoli2021indi} and then exploiting the results derived in \cite{harper2007probability,de2009exchangeable,de2012exchangeability} for CPT.

In Section \ref{sec:QET}, we provided a representation result (Theorem~\ref{th:reprdouble}) for distinguishable particles in QT, where density matrices are quasi-expectations of certain polynomials. In this section, we show that this representation allows us to derive an alternative view of the \textit{second quantisation} for QT and discover the analogous of the Bernstein polynomials \eqref{eq:Berne} for QT.
To achieve this, we will focus only on bosons whose symmetry is similar to that of dice rolls.

In case of two identical bosons, providing this alternative view boils down to deriving the equivalence, under exchangeability, between exchangeable QEOs defined on the following two vector-space of polynomials:
\begin{itemize}
\item $\mathcal{G}=\{({\bf x}_1 \otimes {\bf x}_2)^\dagger G({\bf x}_1 \otimes {\bf x}_2):~~G~~ \text{Hermitian}\}$,
\item $\mathcal{Q}=\{({\bf x} \otimes {\bf x})^\dagger G({\bf x} \otimes {\bf x}):~~G~~ \text{Hermitian}\}$.
\end{itemize}
In the first set, we have two exchangeable particles ${\bf x}_1 ,{\bf x}_2$, while in the second set we have a copy of the same particle ${\bf x}$, resulting in a degree $2$ polynomial.

More generally, consider $\x \in \overline{\mathbb{C}}^{n_x}$ and the following vector space of polynomials:
\begin{equation}
 \label{eq:gambQ}
\mathcal{Q}=\{(\otimes_{i=1}^m\x)^\dagger Q (\otimes_{i=1}^m \x) : Q \text{ is  $n_x^m\times n_x^m$ Hermitian}\}.
\end{equation}
As in Section \ref{sec:QET}, we can define a QEO:
\begin{equation}
 \label{eq:tildeM}
M^p=\widetilde{L}^p((\otimes_{i=1}^m \x)(\otimes_{i=1}^m\x)^\dagger ),
\end{equation}
 satisfying $M^p\succeq0$ and $Tr(M^p)=1$. The superscript $^p$ means `power' and denotes the fact that the monomials in $(\otimes_{i=1}^m \x)(\otimes_{i=1}^m \x)^\dagger $ have power greater than one (compared with $(\otimes_{i=1}^m \x_i)(\otimes_{i=1}^m \x_i)^\dagger $). 

\begin{theorem}[The power-exchangeability representation equivalence for bosons]
\label{th:2}
Let $\x,\x_1,\x_2,\dots,\x_{m}  \in \overline{\mathbb{C}}^{n_z}$ and define
$M=\widetilde{L}((\otimes_{i=1}^m \x_i)(\otimes_{i=1}^m \x_i)^\dagger )$ and
$M^p=\widetilde{L}^p((\otimes_{i=1}^m \x)(\otimes_{i=1}^m \x)^\dagger )$, and
\begin{align}
 \label{eq:states1}
\mathcal{S}_1&=\{M: ~M=\Pi_{Sym}M\Pi_{Sym}, ~M\succeq 0, ~Tr(M)=1\},\\
 \label{eq:states2}
\mathcal{S}_2&=\{M^p:~M^p\succeq0,~Tr(M^p)=1\},
\end{align}
then $\mathcal{S}_1=\mathcal{S}_2$.
\end{theorem}
To explain this result, assume that $n_z=3$ and $m=2$, then the vectors $\x_1\otimes \x_2$, $\Pi_{Sym}(\x_1\otimes \x_2)$
and $\otimes_{i=1}^2 \x$ are respectively equal to:
$$
\begin{tikzpicture}[
    node distance=1mm and 0mm,
    baseline,
          every left delimiter/.style={xshift=.3em},
    every right delimiter/.style={xshift=-.3em}]
    \matrix (M1) [matrix of nodes,{left delimiter=[},{right delimiter=]}]
{
$x_{11}x_{21}$\\[0.25cm]
$x_{11}x_{22}$\\[0.25cm]
$x_{11}x_{23}$\\[0.25cm]
$x_{12}x_{21}$\\[0.25cm]
$x_{12}x_{22}$\\[0.25cm]
$x_{12}x_{23}$\\[0.25cm]
$x_{13}x_{21}$\\[0.25cm]
$x_{13}x_{22}$\\[0.25cm]
$x_{13}x_{23}$\\
};
\end{tikzpicture},~~~
\begin{tikzpicture}[
    node distance=0mm and 0mm,
    baseline,
          every left delimiter/.style={xshift=.3em},
    every right delimiter/.style={xshift=-.3em}]
\matrix (M2) [matrix of nodes,{left delimiter=[},{right delimiter=]}]
{
$x_{11}x_{21}$\\[0.1cm]
$\tfrac{1}{2}(x_{11}x_{22}+x_{12}x_{21})$\\[0.1cm]
$\tfrac{1}{2}(x_{11}x_{23}+x_{13}x_{21})$\\[0.1cm]
$\tfrac{1}{2}(x_{11}x_{22}+x_{12}x_{21})$\\[0.1cm]
$x_{12}x_{22}$\\[0.1cm]
$\tfrac{1}{2}(x_{12}x_{23}+x_{13}x_{22})$\\[0.1cm]
$\tfrac{1}{2}(x_{11}x_{23}+x_{13}x_{21})$\\[0.1cm]
$\tfrac{1}{2}(x_{12}x_{23}+x_{13}x_{22})$\\[0.1cm]
$x_{13}x_{23}$\\
};
\draw[red,very thick] 
        (M2-2-1.north west) -| (M2-2-1.south east) -| (M2-2-1.north west);
\draw[red,very thick] 
        (M2-4-1.north west) -| (M2-4-1.south east) -| (M2-4-1.north west);
\draw[green,very thick] 
        (M2-3-1.north west) -| (M2-3-1.south east) -| (M2-3-1.north west);
\draw[green,very thick] 
        (M2-7-1.north west) -| (M2-7-1.south east) -| (M2-7-1.north west);
\draw[blue,very thick] 
        (M2-6-1.north west) -| (M2-6-1.south east) -| (M2-6-1.north west);
\draw[blue,very thick] 
        (M2-8-1.north west) -| (M2-8-1.south east) -| (M2-8-1.north west);
\end{tikzpicture},~~~
\begin{tikzpicture}[
    node distance=0mm and 0mm,
    baseline,
      every left delimiter/.style={xshift=.3em},
    every right delimiter/.style={xshift=-.3em}]
\matrix (M2) [matrix of nodes,{left delimiter=[},{right delimiter=]}]
{
$x_{1}^2$\\[0.22cm]
$x_{1}x_{2}$\\[0.22cm]
$x_{1}x_{3}$\\[0.22cm]
$x_{2}x_{1}$\\[0.22cm]
$x_{2}^2$\\[0.22cm]
$x_{2}x_{3}$\\[0.22cm]
$x_{3}x_{1}$\\[0.22cm]
$x_{3}x_{2}$\\[0.22cm]
$x_{3}^2$\\
};
\draw[red,very thick] 
        (M2-2-1.north west) -| (M2-2-1.south east) -| (M2-2-1.north west);
\draw[red,very thick] 
        (M2-4-1.north west) -| (M2-4-1.south east) -| (M2-4-1.north west);
\draw[green,very thick] 
        (M2-3-1.north west) -| (M2-3-1.south east) -| (M2-3-1.north west);
\draw[green,very thick] 
        (M2-7-1.north west) -| (M2-7-1.south east) -| (M2-7-1.north west);
\draw[blue,very thick] 
        (M2-6-1.north west) -| (M2-6-1.south east) -| (M2-6-1.north west);
\draw[blue,very thick] 
        (M2-8-1.north west) -| (M2-8-1.south east) -| (M2-8-1.north west);
\end{tikzpicture}
$$
which shows that $\Pi_{Sym}(\x_1\otimes \x_2)$
and $\otimes_{i=1}^2 \x$ have the same symmetries.
Theorem~\ref{th:2} tells us that under indistinguishability, we can swap \textit{exchangeability symmetries} with \textit{power symmetries} in the polynomials. This means that we can express the second quantisation using  the same mathematical objects as in standard QT, but working with the observables defined by the set $\mathcal{Q}$ and with the density matrices $M^p=\widetilde{L}^p((\otimes_{i=1}^m \x)(\otimes_{i=1}^m\x)^\dagger )$. In doing so, the preservation of symmetries is automatically guaranteed. 
Working with $M^p$ results in the second quantisation formalism but expressed in the language of polynomials, see Supplementary \ref{app:2ndquant}.  

We can finally prove the following result.

\begin{corollary}[Representation theorem for probabilities for bosons]
\label{co:gleasonbosons}
Let  $P$ be a probability satisfying \eqref{eq:p1gl}--\eqref{eq:p2gl}  for each  orthogonal basis $(\mathbf{z}_1, \dots, \mathbf{z}_{n_z})$ such that 
$\Pi_{Sym}\mathbf{z}_i=\mathbf{z}_i$. Then  there exists a QEO $\widetilde{L}^p$ as in Theorem~\ref{th:2} such that:
\begin{equation}
P(\mathbf{z}_i)=\widetilde{L}^p((\otimes_{i=1}^m \x)^\dagger  \mathbf{z}_i\mathbf{z}^\dagger_i(\otimes_{i=1}^m \x)).
\end{equation}
\end{corollary}

The monomials in $(\otimes_{i=1}^m \x)(\otimes_{i=1}^m \x)^\dagger$ are the quantum analogs of the Bernstein polynomials \eqref{eq:Berne}, indeed, reasoning about identical bosons can be reduced to reasoning about the polynomials $(\otimes_{i=1}^m \x)^\dagger G(\otimes_{i=1}^m \x)$.
We will show an important consequence of Theorem~\ref{th:2} in the next section.

\section{Convergence to classical reality}
\label{sec:conv}
The above formulation of the second quantisation can be used to show how classical reality emerges in QT for a system of identical bosons due to particle \textit{indistinguishability}. The phenomenon is analogous to the one discussed in Section \ref{sec:defin} for classical dices. It arises when we try to isolate a part of a system from the rest, but the two subsystems remain `paired' due to \textit{exchangeability} symmetries resulting from indistinguishability.

 In what follows, we focus on a system of two distinguishable bosons $\x_1,\y$, although the result is general. The next lemma, proved in \cite[Th.2]{doherty2004complete}, is  key to derive the emergence of classical reality.
\begin{lemma}
\label{lem:SOSwitness}
Consider the polynomial $f([\x_1,\y], [\x_1^\dagger,\y^\dagger])=(\x_1 \otimes \y)^{\dagger} W (\x_1 \otimes \y)$ of complex variables $\x_1 \in \overline{\mathbb{C}}^{n_x},\y\in \overline{\mathbb{C}}^{n_y}$ and $W$ is Hermitian. The following conditions are equivalent.
\begin{enumerate}
    \item $f([\x_1,\y], [\x_1^\dagger,\y^\dagger])>0$ for all   $\x_1 \in \overline{\mathbb{C}}^{n_x},\y\in \overline{\mathbb{C}}^{n_y}$;
    \item there exist positive integers $r$ such that 
    $f([\x_1,\y], [\x_1^\dagger,\y^\dagger]) (\x_1^\dagger \x_1)^{r} \in \Sigma^+_{n_x^{r+1}n_y}$;
\end{enumerate}
where $\Sigma^+_d$ is the closed convex cone of Hermitian sum-of-squares polynomials of dimension $d$. 
\end{lemma}

Notice that Lemma~\ref{lem:SOSwitness} is similar to Lemma~\ref{lem:lemwitndice} for the classical dice.\footnote{The difference is that here we are considering a partial finitely exchangeable setting, meaning that only the $\x$s are exchangeable (not the $\y$). In Lemma~\ref{lem:lemwitndice}, we instead assumed that all the rolls of the dice were exchangeable. This is not a big issue, Lemma~\ref{lem:lemwitndice} can be extended to  the partial exchangeability setting by using the results in \cite{de2016representation}.}  Analogously to the latter, the former lemma explains the emergence of classical reality. 
%We elucidate this observation  
%We show how Lemma~\ref{lem:SOSwitness} explains the emergence of classical reality 
%with an example.

To elucidate this, consider a system composed by two entangled \textit{distinguishable bosons} ${\bf x}_1,\y$. Let $W$ be an entanglement witness  for  ${\bf x}_1,\y$. By definition this means that
$$
f([\x_1,\y], [\x_1^\dagger,\y^\dagger])=(\x_1 \otimes \y)^{\dagger} W (\x_1 \otimes \y)>0,
$$
 for all $ \x_1 \in \overline{\mathbb{C}}^{n_x},\y\in \overline{\mathbb{C}}^{n_y}$.
Assume we consider a system of additional $r$ bosons ${\bf x}_2,{\bf x}_3,\dots,{\bf x}_{r+1}$ such that ${\bf x}_1,{\bf x}_2,\dots,{\bf x}_{r+1}$ are \textit{indistinguishable} (but not with $\y$). 

Consider now the following equalities:
\begin{equation}
    \label{eq:polydecomp}
    \begin{aligned}
0&< (\x_1 \otimes \y)^{\dagger} W (\x_1 \otimes \y)&\\
&=(\x_1 \otimes \y)^{\dagger} W (\x_1 \otimes \y)\prod_{i=2}^{r+1}\x_i^{\dagger}\x_i& \color{darkgray}{(\x_i^{\dagger}\x_i=1)}\\
&=(\otimes_{i=1}^{r+1} \x_i \otimes \y)^{\dagger} (I_{rn_x} \otimes W) (\otimes_{i=1}^{r+1} \x_i  \otimes \y ),&\\
&=(\x_1 \otimes \y)^{\dagger} W (\x_1 \otimes \y)\prod_{i=2}^{r+1}\x_1^{\dagger}\x_1& \color{darkgray}{(\x_1^{\dagger}\x_1=1)}\\
&=(\otimes_{i=1}^{r+1} \x_1 \otimes \y)^{\dagger} (I_{rn_x} \otimes W) (\otimes_{i=1}^{r+1} \x_1  \otimes \y ).&
\end{aligned}
\end{equation}
The above procedure is the analogous of the \textit{degree extension} in \eqref{eq:LPextension}.
%where $\otimes_{i=1}^{r+1} \x_i:=\x_1 \otimes \x_2 \otimes \dots \otimes \x_{r+1}$. 
Lemma~\ref{lem:SOSwitness} states that there exists a positive integer $r$ such that the polynomial $f([\x_1,\y], [\x_1^\dagger,\y^\dagger])(\x_1^\dagger\x_1)^r$ is in the closed convex cone of hermitian SOS polynomials.
%This means we can rewrite the last polynomial in \eqref{eq:polydecomp} as $(\otimes_{i=1}^{r+1} \x_1 \otimes \y)^{\dagger} G (\otimes_{i=1}^{r+1} \x_1  \otimes \y )$, where the Hermitian matrix $G$ is negative semi-definite.

Equivalently, by duality,  for all  matrices
\begin{equation}
    \label{eq:dmb}
    M^p=\widetilde{L}^p\left((\otimes_{i=1}^{r+1}\x_1 \otimes \y)(\otimes_{i=1}^{r+1}\x_1 \otimes \y)^{\dagger}\right),
\end{equation}
we have that $Tr((I_{n_xr} \otimes W)M^p )\geq 0$. By Theorem~\ref{th:2}, this implies that
$$
Tr((I_{n_xr} \otimes W)M^p)=Tr( (I_{n_xr} \otimes W)\Pi^x_{Sym}\rho\Pi^x_{Sym})\geq 0
$$
for all density matrices $\rho$. This shows that entanglement between the $\x,\y$ vanishes at the increase of $r$.

%In other words, we have proven the following. Let $\rho$ be the density matrix for ${\bf x}_1,{\bf x}_2,\dots,{\bf x}_{r+1},\y$ and define
%\begin{align}
%\label{eq:cost}
%\gamma=&\min_{\rho} Tr(W \rho_{xy} )\\
%\nonumber
%&s.t.\\
%\label{eq:symcon}
%&\rho = \Pi^x_{Sym} \rho \Pi^x_{Sym},
%\end{align}
%where $\Pi^x_{Sym}$ is the symmetriser that models the indistinguishability of the $x$s and $\rho_{xy}$ is the reduced density matrix obtained from $\rho$ by (symmetrically) marginalising out all the xs besides one.
%Since  ${\bf x}_1,\y$ are entangled, then $\gamma<0$ for $r=0$.  Lemma~\ref{lem:SOSwitness} proves that exists $r\geq0$ such that $\gamma\geq 0$. 
This means that the marginal quantum system ${\bf x},{\bf y}$ approaches a classical system at the increase of $r$. More precisely, the entanglement between ${\bf x},{\bf y}$ asymptotically disappears at the increase of $r$.

Note that,  \textbf{indistinguishability of the $r+1$ bosons is the key element for the emergence of classical reality} in this setting.
Indeed, if the particles were distinguishable, we could find a density matrix $\rho$ for the composed system of $r+2$ particles such that $Tr((I_{n_xr} \otimes W)\rho)<0$ (for instance, using $\rho=I_{n_xr} \otimes \rho_e$, where $\rho_e$ is the maximum entangled state relative to $W$).

%Lemma~\ref{lem:SOSwitness} states that classical reality emerges at the increase of $r$. 

From Lemma~\ref{lem:SOSwitness}, we can therefore derive  the following result.

\begin{theorem}
\label{th:finalbosons}
Assume we have a two distinguishable particles system $\x,\y$ and additional $r$ bosons that are indistinguishable to $\x$.  We denote the overall composite system of $r+2$ particles as $\mathcal{A}_{r+2}$ and the marginal system  $\x,\y$ as $\mathcal{A}_{r+2|2}$. Then, due to indistinguishability, the subsystem $\mathcal{A}_{r+2|2}$ tends to a classical system as $r$ increases. 
\end{theorem}

\begin{example}
%We show an example of the convergence. 
Consider  the entanglement witness
\begin{equation}
    \label{eq:W}
    W=\left[\begin{matrix}0.25 & 0.0 & 0.0 & -1.0\\0.0 & 2.25 & -1.0 & 0.0\\0.0 & -1.0 & 2.25& 0.0\\-1.0 & 0.0 & 0.0 & 0.25\end{matrix}\right],
\end{equation}
The maximum entangled density matrix is 
$$
\rho_e=\frac{1}{2}\begin{bmatrix}
      1 & 0 & 0 &1\\
      0 & 0 & 0 &0\\
      0 & 0 & 0 &0\\
      1 & 0 & 0 &1\\
     \end{bmatrix},
$$
which satisfies $Tr(W \rho_e)=-0.75<0$ corresponding to the minimum eigenvalue of $W$ (its eigenvalues are $\{-0.75, 1.25,1.25, 3.25\}$). The polynomial
\begin{equation}
\label{eq:sosqt}
\begin{aligned}
&f([\x_1,\y], [\x_1^\dagger,\y^\dagger])=(\x_1 \otimes \y)^{\dagger} W (\x_1 \otimes \y)\\
&=
 2 x_{1} x_1^{\dagger} y_{2} y_2^{\dagger} -  x_{1} x_2^{\dagger} y_{1} y_2^{\dagger} -  x_{1} x_2^{\dagger} y_1^{\dagger} y_{2} -  x_1^{\dagger} x_{2} y_{1} y_2^{\dagger} \\
 &-  x_1^{\dagger} x_{2} y_1^{\dagger} y_{2} + 2 x_{2} x_2^{\dagger} y_{1} y_1^{\dagger}+0.25\geq 0.25>0,
 \end{aligned}
\end{equation}
where $\x_1=[x_1,x_2]^\top$ and $\y=[y_1,y_2]^\top$, is strictly positive. This shows that $W$ is an entanglement witness for $\rho_e$

Any classical expectation of $f$ must satisfy $L(\x_1 \otimes \y)^{\dagger} W (\x_1 \otimes \y))\geq 0.25$, instead 
$\widetilde{L}(\x_1 \otimes \y)^{\dagger} W (\x_1 \otimes \y))=Tr(W \rho_e)=-0.75$ is negative.

Figure \ref{fig:convbosn} reports the value of $\widetilde{L}^p\left((\otimes_{i=1}^{r+1} \x_1 \otimes \y)^{\dagger} (I_{rn_x} \otimes W) (\otimes_{i=1}^{r+1} \x_1  \otimes \y )\right)$ as function of $r$. It shows that entanglement between the $\x,\y$ vanishes at the increase of $r$.

\begin{figure}[htp!]
    \centering
  \includegraphics[width=7cm]{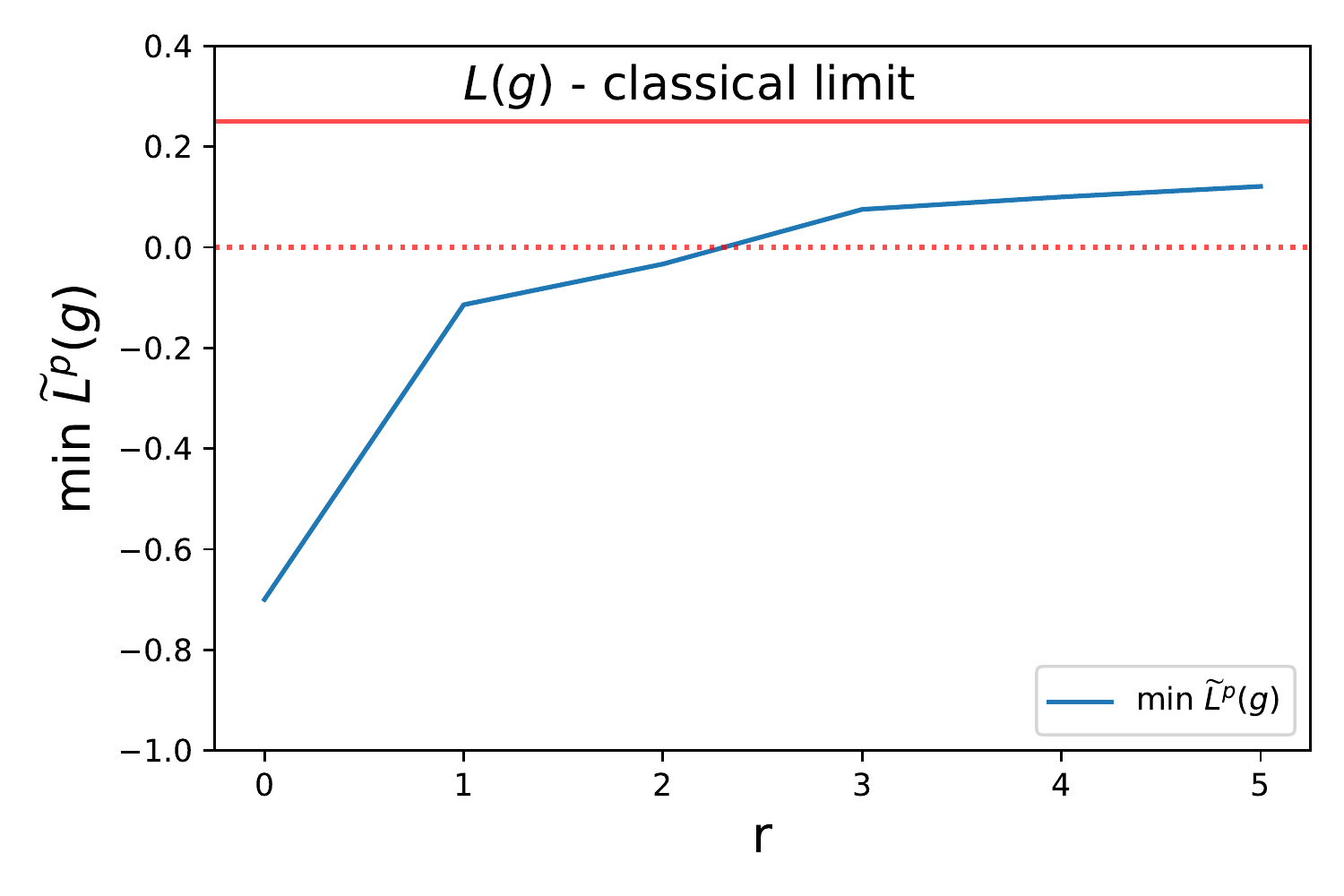}
  \caption{Convergence to classical reality, that is convergence of $\widetilde{L}^p(g)$  to ${L}(g)$ at the increase of  $r$. We stopped at $r=5$ due to computational resources issues (to solve the corresponding semi-definite programming problem).}
  \label{fig:convbosn}
\end{figure}
\end{example}

\begin{figure}[!htp]
\centering
\includegraphics[width=10cm]{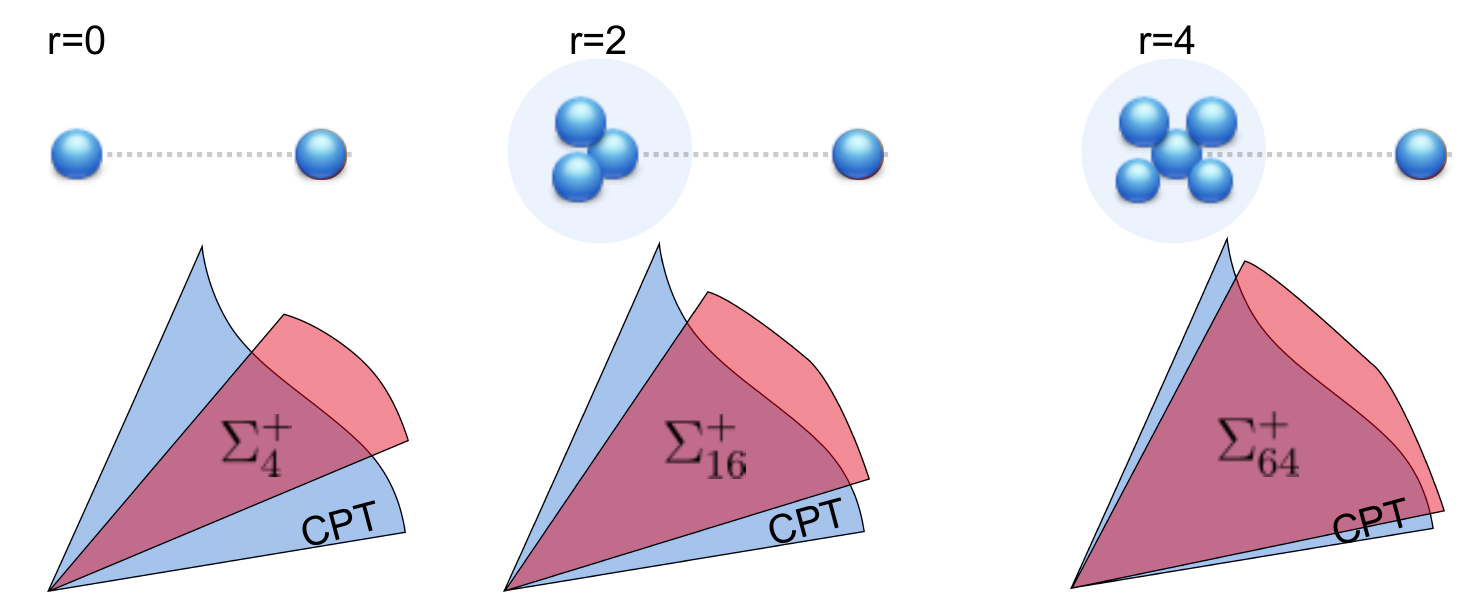}
\caption{Pictorial representation of the convergence of the SOS Hermitian cone of polynomials to the cone of nonnegative polynomials (CPT) for $r+1$ identical bosons.}
\label{fig:conesqt}
\end{figure}

Similar to quantum decoherence, the exact convergence only happens for  $r \rightarrow \infty$, but the difference between CPT and QT becomes quickly small as $r$ increases. In fact, at the increase of the number $r$ of identical bosons, the cone of Hermitian sum-of-squares polynomials converges to the cone of nonnegative polynomials of the form $(\otimes_{i=1}^{r+1} \x_1 \otimes \y)^{\dagger} G (\otimes_{i=1}^{r+1} \x_1  \otimes \y )$. A pictorial representation is given in Figure \ref{fig:conesqt}.

\section{Discussion}
In this work we have shown how the formulation of QT as  a quasi-expectation operator (QEO) provides a direct interpretation of density matrices as moment matrices. Moreover,  using  the classical statistical concept of exchangeability, this formalism allows us to directly derive the symmetrisation postulate for identical particles and, also, derive a novel representation for the second quantisation for bosons in terms of $m$-degree polynomials ($m$ being the number of indistinguishable particles).

By exposing the connection between indistinguishable particles (bosons) and finitely exchangeable random variables (rolls of a classical dice), we were able to show how classical reality emerges due to indistinguishability in a system of identical bosons. This was achieved by providing a \textit{representation theorem} {\`a} la de Finetti for bosons.

The keystone of these results is the derivation of QT as a QEO.
As we have seen, a QEO is defined by two properties: (i) linearity; (ii) lower bound of the minimum of its argument (property \eqref{eq:Astar}). Given a linear operator  $\widetilde{L}$, we call \emph{validation problem} the problem of deciding whether $\widetilde{L}$ satisfies properties \eqref{eq:Astar}. If this validation problem can be solved in \textbf{P}-time, we say that QEO is tractable. 

%\paragraph{Linearity.} QEO and expectation operators are both linear, why is linearity a reasonable assumption? @Marco

%\paragraph{Symmetries versus tractability}
%In the previous sections, we have presented two cases in which a \textit{representation theorem} {\`a} la de Finetti relates a classical probability mass function satisfying certain symmetries to a QEO over a vector space of polynomials (basis functions). 
%:$$
%\textit{symmetric probability}= \widetilde{L}(\textit{polynomials}).
%$$
%In both cases, the validation problem associated to the QEO is tractable: using linear programming in case of the classical dice, and using SemiDefinite Programs (SDP) for QT. The immediate question is how symmetries and tractability relates. Indeed, both CPT and QT, given our framework, satisfies the same symmetries constraints. However CPT satisfies additional, notice that Do symmetries explain tractability, or is the other BLABLABLA

%\paragraph{Connection to SDP.}
 The validation problem associated to the QEO is tractable for QT by %using : using linear programming in case of the classical dice, and 
 using SemiDefinite Programming (SDP). 
 The framework of SDP  is ubiquitous in quantum information. It is for instance employed as a tool to assess when a density matrix is entangled. In this respect, Doherty-Parrilo-Spedalieri (DPS) \cite{doherty2002distinguishing,doherty2004complete} introduced a hierarchy of SDP relaxations to the set of separable states, which is defined in terms of so-called \textit{state extension} (what we called \textit{degree extension}) whose convergence follows by Lemma~\ref{lem:SOSwitness}. The convergence of the DPS hierarchy was proven in \cite{doherty2004complete} by using the quantum de Finetti Theorem \cite{Caves02}.
In this paper, we have shown that this result is more than a simple optimisation trick to prove the separability of an entanglement matrix. By exploiting our formulation of QT as a QEO, we have shown that the DPS hierarchy really exists in Nature and is responsible for the emergence of classical reality for identical bosons. 

\paragraph{Connection to the quantum de Finetti Theorem.} 
Consider a quantum experiment which aims to measure the state of a distinguishable particle by $N$ repeated measurements. Consider also an experimenter who judges the collection of the $N$ measurements (the device's outputs) to have an overall quantum state $\rho^{(N)}$. The experimenter will also judge any permutation of those outputs to have
the same quantum state $\rho^{(N)}$ (for any $N$).\footnote{There is an additional consistency condition that any $\rho^{(N)}$ can be derived by $\rho^{(N+1)}$.} The quantum de Finetti Theorem \cite{Caves02} states that  $\rho^{(N)}$ is an infinitely exchangeable sequence of states if and only if it can be written as
\begin{equation}
\label{eq:qdefinth}
    \rho^{(N)} = \int \left(\otimes_{i=1}^N \rho\right) du(\rho),
\end{equation}
where $u(\rho)$ is a probability distribution over the density operator $\rho$. By using the interpretation of $\rho$ as moment matrix, then   $u(\rho)$ can be understood as a probability distribution over a moment matrix (similar to the Wishart distribution).

By using the results of this paper, we can then prove:

\begin{proposition}
\label{prop:qdefinequiv}
$\rho^{(N)}$ is an infinitely exchangeable sequence of states if and only if it can be written as
\begin{equation}
\label{eq:qdefinth1}
    \rho^{(N)} = \int \left(\otimes_{i=1}^N \x\x^{\dagger}\right) dp(\x)=L(\otimes_{i=1}^N\x\x^{\dagger})=L((\otimes_{i=1}^N\x)(\otimes_{i=1}^N \x)^{\dagger}),
\end{equation}
for some probability measure $p$.
\end{proposition}
This holds in the infinitely exchangeable case. As for the classical dice, for a  finitely exchangeable sequence (only $m$ repetitions), the quantum de Finetti theorem does not hold. In this case, by exploitng the results of the previous sections, we can derive that 
\begin{equation}
\label{eq:qdefinth2}
    \rho^{(m)} = \int \left(\otimes_{i=1}^m \x\x^{\dagger}\right) d\nu(\x)=\widetilde{L}((\otimes_{i=1}^N\x)(\otimes_{i=1}^N \x)^{\dagger}),
\end{equation}
where $\nu$ is a signed-measure. Indeed, $\widetilde{L}((\otimes_{i=1}^N\x)(\otimes_{i=1}^N \x)^{\dagger})$ is what we called $M^p$. By exploiting the  power-exchangeability representation equivalence for bosons (Theorem \ref{th:2}) and the equivalence in Proposition \ref{prop:qdefinequiv}, we can then understand how the symmetries underling the quantum de Finetti theorem are related to the symmetries of identical bosons.

\paragraph{Related approaches for the emergence of classical reality.} It is worth to connect the  results derived in this paper for identical bosons to two approaches that studied the emergence of classical reality.
\begin{itemize}
    \item Using the quantum de Finetti theorem, \cite{chiribella2006quantum} considered the problem of a quantum channel that equally distributes information among $m$ users, showing that for large $m$ any such channel can be efficiently approximated by a classical one.  
    \item Decoherence \cite{zurek2003decoherence} provides a possible explanation for the quantum-to-classical transition by appealing to the immersion of nearly all physical systems in their environment.
 This typically leads to the selection of persistent pointer states, while superpositions of such pointers states are suppressed. Pointer states (and their  convex combinations) become natural candidates for classical states. However, decoherence does not explain how information about the pointer states reaches the observers, and how such information becomes objective, that is, agreed upon by several observers. Quantum Darwinism tries to overcome this issue by interpreting pointer observables as  information about a physical system that the environment selects and proliferates among $m$ observers. Using the quantum de Finetti theorem, \cite{brandao2015generic} proved that classical reality emerges at the increase of $m$. Indeed, the setting we used in the last example paper is similar to the one described in \cite{ccakmak2021quantum}: two-qubit system coupled to an N-qubit system. 
\end{itemize}
 Both these two results could be derived directly from \cite{doherty2004complete} by literally interpreting \textit{state extension} (\textit{degree extension})  as a cloning procedure.
 In this work, we have shown that the emergence of classical reality  is also due to indistinguishability of identical bosons. The key point is the equivalence between exchangeability symmetries and power symmetries (degree extension). 
 
As future work, we plan to extend this result to fermions. In this case, we must face the issue that the hierarchy (the degree $r$) cannot grow arbitrarily large due to the anti-symmetric behaviour of fermions. However, for systems with large degrees of freedom, we expect a similar convergence to also hold for fermions.

\section*{References}
\bibliographystyle{ieeetr}
\bibliography{biblio}% Produces the bibliography via BibTeX.

\appendix

\section{Connection with the second quantisation}
\label{app:2ndquant}
Assume we have $m$ bosons, the one-particle states form an orthonormal basis $span(\ket{v_1},\dots,\ket{v_m})$ of $V=\overline{\mathbb{C}}^n$ and, therefore, the joint state is in $V^{\otimes m}$ with basis
$$
\ket{v_i}_{(1)}\otimes \dots \otimes \ket{v_p}_{(m)}.
$$
By applying $\Pi_{Sym}$ to the full set of states in  $V^{\otimes m}$ we obtain $\text{Sym}^m V$, that is the symmetric vector space of the $m$ particles. To distinguish states in $V^{\otimes m}$ that are mapped into the same element in $\text{Sym}^m V$ by $\Pi_{Sym}$, we define the \textit{occupation number}. An occupation number is an integer $n_i \geq 0$ associated with each vector in $V$:
$$
\underset{n_1}{\ket{v_1}},\underset{n_2}{\ket{v_2}},\dots,\underset{n_m}{\ket{v_m}}
$$
where each $n_i$ tells us the number of times that $\ket{v_i}$ appears in the chosen basis state in $V^{\otimes m}$. Two basis states in $V^{\otimes m}$ with the same occupation numbers will be mapped into the same element in $\text{Sym}^m V$ and, therefore they form a class of equivalence in $V^{\otimes m}$ which we denote as
$$
\ket{n_1,n_2,\dots,n_m},
$$
and these basis states forms a basis in $\text{Sym}^m V$.

Equivalently, we can interpret the vector $\ket{n_1,n_2,\dots,n_m}$ as the exponents of the monomials $\otimes_{i=1}^m \x \in \bar{\mathbb{C}}^{n}$. For instance, assume that $n=3$ and $m=2$, $\ket{v_i}={\bf e}_i$ (the canonical basis in $\mathbb{R}^3$) then  $\x_1\otimes \x_2$, $\ket{2,0,0}$, $\ket{1,1,0}$
and $ \ket{1,0,1}$ are, respectively, equal to:
$$
\begin{bmatrix}
x_{11}x_{21}\\
x_{11}x_{22}\\
x_{11}x_{23}\\
x_{12}x_{21}\\
x_{12}x_{22}\\
x_{12}x_{23}\\
x_{13}x_{21}\\
x_{13}x_{22}\\
x_{13}x_{23}\\
\end{bmatrix},~~
\left\{\begin{bmatrix}
1\\
0\\
0\\
0\\
0\\
0\\
0\\
0\\
0\\
\end{bmatrix}\right\},~~~\left\{\begin{bmatrix}
0\\
1\\
0\\
0\\
0\\
0\\
0\\
0\\
0\\
\end{bmatrix},\begin{bmatrix}
0\\
0\\
0\\
1\\
0\\
0\\
0\\
0\\
0\\
\end{bmatrix}\right\},~~\left\{\begin{bmatrix}
0\\
0\\
1\\
0\\
0\\
0\\
0\\
0\\
0\\
\end{bmatrix},\begin{bmatrix}
0\\
0\\
0\\
0\\
0\\
0\\
1\\
0\\
0\\
\end{bmatrix}\right\}
$$
where the curling bracket denotes an equivalence class.
These equivalence classes correspond to the monomials $x_1^2$, $x_1x_2$ and, respectively, $x_1x_3$ of the vector $\otimes_{i=1}^2 \x$, which have degree $(2,0,0)$, $(1,1,0)$ and $(1,0,1)$ w.r.t.\ the variables $ \x=[x_1,x_2,x_3]^\dagger$.

\section{Proofs}
\label{sec:proofs}
\paragraph{Proposition~\ref{prop:locald}}
We exploit  the \textit{mixed-product property} of the Kronecker product.
 $$
 \begin{aligned}
&g(\x,\x^\dagger)h(\y,\y^\dagger)={\bf x}^\dagger F {\bf x}{\bf y}^\dagger H {\bf y}=  ({\bf x}^\dagger F {\bf x})\otimes ({\bf y}^\dagger H {\bf y})\\
    &=  ({\bf x}^\dagger \otimes {\bf y}^\dagger)(G{\bf x} \otimes H {\bf y})=  ({\bf x} \otimes {\bf y})^\dagger( F {\bf x} \otimes  H {\bf y})\\
    &=  ({\bf x} \otimes {\bf y})^\dagger( F  \otimes  H)({\bf x} \otimes {\bf y}).\\
 \end{aligned}
 $$
The second part is a known property of the space of Hermitian matrices, which is usually called `tomographic locality' or `local discriminability', see for instance \cite{dariano_chiribella_perinotti_2017}.

\paragraph{Theorem~\ref{th:reprsingle} and Theorem~\ref{th:reprdouble}}
It follows directly from the results in \cite[Appendix B2]{Benavoli2021f}.

\paragraph{Corollary~\ref{co:gleason}}
Any projection matrix $\Pi_i$ can be written as $\mathbf{z}_i\mathbf{z}_i^\dagger$ for a $\mathbf{z}_i \in \overline{\mathbb{C}}^{n_z}$.

\paragraph{Theorem~\ref{th:classidiceconvergebnce}}
 Theorem~\ref{th:classidiceconvergebnce} is a particular case of the so-called Krivine-Vasilescu's nonnegativity criterion, obtained  considering as possibility space  $\mathbb{K}$ the probability simplex: 
\begin{equation}
\label{eq:OmegaconstrStheta}
 \mathbb{K}=\left\{\theta \in \reals^5:  ~ \theta_j\geq0, ~~1-\sum_{j=1}^5 \theta_j\geq 0 \right\}.
\end{equation}
It can be proven as a special case of  \cite[Th.\ 5.11]{lasserre2009moments} with $\widehat{g}^{\boldsymbol \alpha}$ corresponding to $\theta_1^{n_1}\theta_2^{n_2}\cdots(1-\theta_1-\dots-\theta_5)^{n_6}$.
The  proof of  \cite[Th.\ 5.11]{lasserre2009moments} uses  Lemma~\ref{lem:lemwitndice} and duality.

\paragraph{Lemma~\ref{lem:lemwitndice}}
It can be proven as a special case of  \cite[Th.\ 2.24]{lasserre2009moments} with $g_1^{ \alpha_1}g_2^{ \alpha_2}\dots g_m^{ \alpha_m}$ corresponding to $\theta_1^{n_1}\theta_2^{n_2}\cdots(1-\theta_1-\dots-\theta_5)^{n_6}$ and $\mathbb{K}$ as in \eqref{eq:OmegaconstrStheta}.

\paragraph{Theorem~\ref{th:2}}
Given both the matrices $M,M^p$ are PSD with trace one, we must only prove that 
$\widetilde{L}^p((\otimes_{i=1}^m \x)(\otimes_{i=1}^m \x)^\dagger )$ and $\widetilde{L}((\x_1\otimes \dots \otimes \x_m)(\x_1\otimes \dots \otimes\x_m)^\dagger))$ have the same symmetries. This follows by the fact: (i) we can consider $\x$ and $\x^\dagger$ as two different variables; (ii) $\Pi_{Sym}(\x_1\otimes \dots \otimes \x_m)$ and $\otimes_{i=1}^m \x$ have the same symmetries, which follows by the definition of the symmetriser operator $\Pi_{Sym}$.

\paragraph{
Lemma~\ref{lem:SOSwitness}}
This result follows from a result derived by Doherty, Parrillo and Spedalieri (DPS) \cite[Th.2]{doherty2004complete}, which generalises the results derived by Quillen \cite{quillen1968representation} and Catlin and D’Angelo \cite{Catlin1996}.

\paragraph{Theorem~\ref{th:finalbosons}}
It follows from Lemma~\ref{lem:SOSwitness} and duality.

\paragraph{Proposition~\ref{prop:qdefinequiv}}
The density matrix $ \rho^{(N)} $ is  written as
$$
 \rho^{(N)} = \int \left(\otimes_{i=1}^N \rho\right) du(\rho)
$$
which means that $ \rho^{(N)} $ is a separable density matrix. This also means that $ \rho^{(N)} $ is a truncated moment matrix and, therefore, it can be written as 
$$
 \rho^{(N)} = \int \left(\otimes_{i=1}^N \x\x^{\dagger}\right) dv(\x).
$$

\end{document}